
\input harvmac.tex

\let\bar=\overline
\def\nl{\hfill\break}
\let\<=\langle
\let\>=\rangle
\let\e=\epsilon
\let\tht=\theta
\let\t=\theta

\let\r=\rho
\let\l=\lambda
\let\p=\prime
\def\ind{Tr $(-1)^F \ F \ e^{-\beta \rm H }$}

\def\Zn{${\bf Z}_n$}
\def\F{$\Tr F(-1)^{F}e^{-\beta H}$}
\noblackbox
\pretolerance=750
\lref\fs{P. Fendley and H. Saleur, Boston and Yale preprint  BUHEP-92-15,
YCTP-P13-1992}
\lref\rLL{L. Landau and E. Lifshitz, {\it Statistical Physics}
(Pergamon, New York, 1980), section 77}
\lref\LGSol{P. Fendley, S.D. Mathur, C. Vafa and N.P. Warner,
Phys. Lett. B243 (1990) 257}
\lref\pk{P. Fendley and K. Intriligator, Nucl. Phys. B372 (1992) 533.}
\lref\rPF{P. Fendley, Boston preprint
BUHEP-91-16, to appear in Nucl Phys. B}
\lref\pkii{P. Fendley and K. Intriligator, Scattering and thermodynamics
in integrable $N$=2 theories, BUHEP-92-5,HUTP-91/A067, to appear in Nucl.
Phys. B}
\lref\rJR{R. Jackiw and C. Rebbi, Phys. Rev. D13 (1976) 3398}
\lref\rGW{J. Goldstone and F. Wilczek, Phys. Rev. Lett. 47 (1981)
986}
\lref\rZamc{A.B. Zamolodchikov, JETP Lett. 43 (1986) 730}
\lref\witti{E. Witten, Nucl. Phys. B202 (1982) 253}
\lref\nonin{C. Imbimbo and S. Mukhi, Nucl. Phys. B247 (1984) 471}
\lref\cab{C. Callias, Commun. Math. Phys. 62 (1978)213; \nl
R. Bott and R. Seeley, Commun.  Math. Phys. 62 (1978) 235}
\lref\tlg{C. Vafa, Mod. Phys. Lett. A6 (1991) 337}
\lref\cec{S. Cecotti, Int. J. Mod. Phys. A6 (1991) 1749; \nl
S. Cecotti, Nucl. Phys. B355
(1991) 755}
\lref\Witd{E. Witten, Commun. Math. Phys. 117 (1988) 353}
\lref\cpg{S. Cecotti, L. Girardello and A. Pasquinucci,
Int. J. Mod. Phys. A6 (1991) 2427}
\lref\rLG{A.B. Zamolodchikov Sov. J. Nucl. Phys. 44 (1986)
529;\nl D. Kastor, E. Martinec and S. Shenker, Nucl. Phys. B316 (1989) 590}
\lref\rVW{E. Martinec, Phys. Lett. 217B (1989) 431;\nl
C. Vafa and N.P. Warner, Phys. Lett. 218B (1989) 51}
\lref\rWO{E. Witten and D. Olive, Phys. Lett. 78B (1978) 97}
\lref\toda{P. Fendley, W. Lerche, S.D. Mathur and N.P. Warner,
Nucl. Phys. B348 (1991) 66}
\lref\rNS{A.J. Niemi and G.W. Semenoff, Phys. Rep. 135 (1986) 99}
\lref\rZandZ{A.B. Zamolodchikov and Al.B. Zamolodchikov, Ann.
Phys. 120 (1980) 253}
\lref\rAZi{A.B. Zamolodchikov, Adv. Stud. Pure Math. 19 (1989) 1}
\lref\rKMi{T.R. Klassen and E. Melzer, Nucl. Phys. B350 (1990) 635}
\lref\tbaref{Al.B. Zamolodchikov, Nucl. Phys. B342 (1990) 695}
\lref\rAlZii{Al.B. Zamolodchikov, Nucl. Phys. B358 (1991) 497}
\lref\rBL{D. Bernard and A. LeClair, Phys. Lett. 247B (1990) 309}
\lref\rGep{D. Gepner,  Fusion rings and geometry, preprint
NSF-ITP-90-184}
\lref\lvw{W. Lerche, C. Vafa, N. Warner, Nucl.  Phys. B324 (1989)
427}
\lref\dvv{R. Dijkgraaf, E. Verlinde, and H. Verlinde, Nucl. Phys.
B352 (1991) 59}
\lref\rken{K. Intriligator, Mod. Phys. Lett. A6 (1991) 3543}
\lref\cv{S. Cecotti and C. Vafa, Nucl. Phys. B367 (1991) 359}
\lref\rNar{K.S. Narain, Nucl. Phys. B243 (1984) 131}
\lref\witN{E. Witten, Nucl. Phys. B149 (1979) 285}
\lref\witten{E. Witten, Nucl. Phys. B340 (1990) 281}
\lref\rGepii{D. Gepner and A. Schwimmer,  Weizmann preprint
WIS-92/34/March-Ph,\nl hepth\#9204020}
\lref\Bord{M. Bourdeau, E. J. Mlawer, H. Riggs, and H.J. Schnitzer,
Brandeis preprint BRX-TH-327 (1991)}
\lref\intcpnrefs{E. Abdalla, M.C. Abdalla, and M. Gomes, Phys.
Rev. D25 (1982) 425; Phys. Rev. D27 (1983) 825}
\lref\rCVii{S. Cecotti and C. Vafa, Phys. Rev. Lett. 68 (1992) 903}
\lref\RSOS{A.B. Zamolodchikov, Landau Institute preprint,
September 1989}
\lref\rABF{G.E. Andrews, R.J. Baxter and P.J. Forrester, J. Stat.
Phys. 35 (1984) 193}
\lref\Kob{K. Kobayashi and T. Uematsu, preprint KUCP-41
(1991), hepth\#9112043}
\lref\bill{W. Leaf-Herrmann, Harvard preprint HUTP-91-A061,
and to appear.}
\lref\nw{D. Nemeschansky and N. Warner, USC preprint USC-91-031}
\lref\rKzS{Y. Kazama and H. Suzuki, Phys. Lett. 216B (1989) 112;
Nucl. Phys. B321 (1989) 232}
\lref\rAlZiv{Al.B. Zamolodchikov, Resonance factorized
scattering and roaming trajectories, Ecole Normale preprint ENS-LPS-355}
\lref\eg{T.Eguchi and S.K. Yang, Mod. Phys. Lett. A5 (1990) 1693}
\lref\susy{S.J. Gates {\it et al},
{\it Superspace: One Thousand and One Lessons in
Supersymmetry} (Benjamin-Cummings, Reading, 1983) }
\lref\dub{B. Dubrovin, Geometry and integrability of
topological-antitopological fusion,
Napoli preprint INFN-8/92-DSF}
\lref\mathr{M.Reed and B. Simon, {\it Methods of Modern Mathematical
Physics}, Academic Press (NY) 1979}
\lref\AlFr{L. Alvarez-Gaume and D. Freedman, Commun. Math. Phys. 80 (1981) 443}
\lref\sqm{R. Kaul and R. Rajaraman, Phys. Lett. 131B (1983) 357}
\lref\wu{B. McCoy, J. Tracy, and T.T. Wu, J. Math. Phys. 18 (1977) 1058}
\lref\lnw{A. LeClair, D. Nemeschansky, and N. Warner, to appear}
\lref\rMW{P. Mathieu and M. Walton, Phys. Lett. 254B (1991) 106.}

\Title{\vbox{\baselineskip12pt\hbox{HUTP-92/A021}\hbox{SISSA 68/92/EP}
\hbox{BUHEP-92-14}}}
{\vbox{\centerline{ A New Supersymmetric Index }}}
\vglue .5cm
\centerline{Sergio Cecotti$^\dagger$, Paul Fendley$^\star$, Ken
Intriligator$^\infty$ and Cumrun Vafa$^\infty$}
\vglue .5cm
\centerline{$^\dagger$ International School for Advanced Studies, SISSA-ISAS}
\centerline{Trieste and I.N.F.N., sez. di Trieste, Trieste, Italy}
\vglue .3cm
\centerline{$^\star $Department of Physics, Boston University}
\centerline{590 Commonwealth Avenue, Boston, MA 02215, USA}
\vglue .3cm
\centerline{$^\infty $Lyman Laboratory of Physics}
\centerline{Harvard University, Cambridge, MA 02138, USA}
\vglue 1cm

We show that ${\rm Tr}(-1)^F F e^{-\beta H}$ is an index for $N$=2
supersymmetric theories in two dimensions, in the sense that it is independent
of almost all deformations of the theory.  This index is related to the
geometry of the vacua (Berry's curvature) and satisfies an exact differential
equation as a function of $\beta$.  For integrable theories we can also
compute the index thermodynamically, using the exact $S$-matrix.  The
equivalence of these two results implies a highly non-trivial equivalence of a
set of coupled integral equations with these differential equations, among
them Painleve III and the affine Toda equations.
\Date{4/92}
\newsec{Introduction}
There has been much progress in understanding supersymmetric quantum field
theories in the last decade.  Supersymmetry turns out to be a strong symmetry
principle which allows one to get a firm grip on certain aspects of these
theories.  For example, Witten's index ${\rm Tr}(-1)^F e^{-\beta H}$
\witti\  is an effective tool in addressing questions of
supersymmetry breaking.  It is natural to ask if there are other `index-like'
objects which can be computed exactly and provide further insight into the
structure of supersymmetric theories.  The aim of this paper is to show in
two-dimensional $N$=2 supersymmetric theories, there is such an object: ${\rm
Tr}(-1)^F F e^{-\beta H}$.  We call this an index because it is independent of
almost all deformations of the action.  It, however, {\it does depend} on a
finite set of (relevant or marginal) perturbations in a way which
can be computed exactly.

 Supersymmetric theories in two dimensions are among the simplest quantum
field theories.  Two-dimensional conformal theories with $N$=2 supersymmetry
can be used to construct string vacua and have thus been studied extensively
recently.  All $N$=2 theories in two dimensions, whether or not they are
conformal, have a set of observables, the (supersymmetric) chiral fields,
which form a ring under operator product.  This is called the chiral ring
\lvw\ (for a review see \cv ).
This ring can be computed {\it exactly} using the techniques of topological
field theories \witten\ (see also \refs{\eg , \tlg} ) as all the $N$=2
theories have a topological counterpart (called the `twisted version').  The
study of chiral rings turn out to be a very powerful tool in unravelling the
geometry of the vacua of the supersymmetric theory.  In particular by joining
the topological and anti-topological versions of $N$=2 theories, one can
derive (integrable) differential equations ($tt^*$ equations) to compute the
Berry's curvature for the vacuum bundle of the supersymmetric theory as one
perturbs the $N$=2 theory \cv\ (see also \refs{\cpg ,\cec } ).  It was
observed in \cv\ that the solutions of these equations resemble a kind of
partition function for kinks of the theory.  This, however, remained a
somewhat mysterious connection to be explained.  In this paper we will see
that these computations are related to the new index ${\rm Tr}(-1)^F F
e^{-\beta H}$ which encodes aspects of the spectrum and the interactions of
the kinks.  In particular, the $tt^*$ equations
provide an exact differential equation in
$\beta$ for the new index for any $N$=2 theory.

 These somewhat formal derivations can be checked very explicitly in many
special cases.  In particular when the $N$=2 theory is integrable, the
existence of infinitely-many conserved charges allows one to construct the
$S$-matrix (more or less uniquely)\rZandZ .  In such cases, one can use the
exact $S$-matrix to find integral equations for the non-perturbative partition
function ${\rm Tr}\ e^{-\beta H}$.  This powerful method is known as the
thermodynamic Bethe ansatz (TBA)\tbaref . In particular, the TBA analysis for
a large class of $N$=2 integrable theories in two dimensions was carried out
in \refs{\pk,\pkii}, confirming the conjectured $S$-matrices as in particular
reproducing the correct central charges in the UV limit.  One can extend the
usual TBA analysis by allowing arbitrary chemical potentials, and in
particular one can compute objects such as ${\rm Tr}e^{i\alpha F}e^{-\beta
H}$.  This allows us, as a special case, to compute ${\rm Tr}(-1)^F F
e^{-\beta H}$ in these theories in terms of integral equations.

Thus for integrable theories we seem to have two inequivalent methods to
compute the new index: one in terms of {\it differential equations}
characterizing the geometry of the vacuum bundle, and the other in terms of
{\it coupled integral equations} coming from TBA.  It is a highly non-trivial
check on all these ideas that the {\it solutions} to these equations are the
same.  We have checked this using numerical solutions to both systems of
equations.  Due to the non-linearity of our differential equation and
complexity of the coupled integral equations, we have not been able to show
directly (i.e., analytically) that these are the same.  In fact, turning
things around, physics has predicted a surprising equivalence
between coupled integral equations and certain differential equations (such as
radial affine toda equations), a result which is yet to be proven
mathematically!

The organization of this paper is as follows:

In section 2 we introduce the new index, and discuss in what sense it is an
index (i.e., we see that it is independent of $D-$term perturbations).  The
derivation of this result in this section is very simple but unfortunately
requires a certain formal manipulation which is not always easy to rigorously
justify.  In section 3 we discuss the geometry of vacua and review the results
of \cv .  Here we show how to rephrase our new index as a computation in the
geometry of vacua. In particular we show why our index depends only on
$F$-terms, thus giving a more rigorous derivation of the results of section 2.
Moreover, this allows us to effectively compute the new index in terms
of solutions of certain non-linear differential equations.

In section 4 we discuss the infra-red expansion of the index.  We show in
particular that at least the leading term (the one-particle contribution) and
the next leading term (the two-particle contribution) are universal.  This
means that they just depend on the mass and the central term of the
supersymmetry algebra and the allowed soliton configurations.  In section 5 we
review briefly the results of \pk\ and discuss how the new index can be
computed for integrable theories using the TBA.  In section 6 we consider a
number of examples including $N$=2 sine-Gordon and minimal $N$=2 theories
perturbed by least and most relevant perturbations.  We write down the
differential equations and the integral equations which are presumably
equivalent.  We explicitly check this for some of the examples numerically.
Moreover in this section we use the TBA to compute the more general object
${\rm Tr}(-1)^F F^l e^{-\beta H}$ and show that, for $l>1$ it is {\it
not} an index and it does depend on the choice of D-terms, as expected.

In section 7 we present our conclusions.  Finally in appendix A the $tt^*$
equations are rederived, in a quick but somewhat non-rigorous way in the same
spirit as the arguments in section 2.

\newsec{ \ind }

In this section we discuss the existence of a new supersymmetric `index' for
$N$=2 supersymmetric quantum field theories in two dimensions.  Our emphasis in
this section is just on formulating what this index is; in the following
sections we show how it may be computed.

Let us start with Witten's index $\rm{Tr} (-1)^F e^{-\beta \rm H}$ \witti.
This index is defined for $N\geq 1$ supersymmetric theories in any dimension.
It is an index because it is independent of finite perturbations of the
theory, provided the space is compact and does not break supersymmetry (e.g.,
a $d$-dimensional torus).  The idea is simply that there are two types of
states in the Hilbert space: states which come in pairs $|s \rangle ,
Q|s\rangle $ where $Q$ is the supersymmetry charge with $Q^2=H$, and states
which come isolated, i.e., the ones which are annihilated by $Q$ and are
ground states of the theory with $H=0$.  The pairs, which necessarily have
$H\not= 0$, do not contribute to the Witten index as they have opposite
$(-1)^F$. This follows from the fact that $\{ (-1)^F,Q\} = 0$.
 Therefore this index simply counts the ground states of the
theory weighted with $\pm 1$, depending on the parity of $(-1)^F$.  Any
finite perturbation of the theory does not change this index: if massive
states become ground states they must do so in pairs, so one adds a $+1$ and a
$-1$ to the index. Similarly, the only way a ground state can become massive
is for ground states with opposite $(-1)^F$ to pair up, so again the net
contribution to the index is zero.  This index has been a powerful object in
probing questions of supersymmetry breaking in supersymmetric theories.

It is well known that the above argument does not apply for non-compact
spaces.  Consider a Hilbert space based on $R^d$.  The above argument breaks
down in this case because the eigenvalues of $H$ are typically continuous.  In
particular it may not be true that the density of states for $|s\rangle $ and
$Q|s\rangle $ are equal.  The contribution to the index may be written as
\eqn\wittbeta{(-1)^{f}\int dE (g_+(E)-g_-(E))e^{-\beta E},}
where $g_{\pm}(E)$ are the density of states distributions for $|s\>$ and $Q
|s\>$ and these states contribute $\pm (-1)^f$ to the index.  If $g_+-g_-$ is
nonzero for $E \neq 0$, the contribution of massive states to the index does
not vanish and thus as we change the parameters in the theory the index
changes.  In particular it does depend on $\beta $.  Examples of this
phenomena have been found in some simple quantum-mechanical systems with $N$=1
supersymmetry \nonin\ where it can be computed exactly using the
Callias-Bott-Seeley index theorem \cab .

Let us consider this situation for the $N$=2 supersymmetric theories in $d=2$
where we take space to be the real line.  The $N$=2 supersymmetry algebra
on the real line can be written as
$$Q^{+2}=Q^{-2}=\bar Q^{+2}=\bar Q^{-2}=\{ Q^+,\bar Q^-\} =\{ Q^-,
\bar Q^+\}=0$$
$$[F,Q^{\pm}]=\pm Q^{\pm} \quad [F,\bar Q^{\pm}]=\mp\bar Q^{\pm}$$
$$[Q^\pm,H_{L,R}]=[\bar Q^\pm ,H_{L,R}]=[F,H_{L,R}]=0$$
$$\{ Q^+, Q^-\}=H_L \qquad \{ \bar Q^+, \bar Q^- \}=H_R$$
\eqn\sual{\{ Q^+, \bar Q^+ \} =\Delta \qquad \{ Q^-, \bar Q^- \}=
\bar\Delta }
where $H_{L,R}=H\pm P$, and $\Delta$ is a $c-$number which is the central term
of the supersymmetry algebra. The fermion number $F$ is the charge
corresponding to the global $O(2)$ symmetry
of $N$=2 supersymmetric theories.  We also
have $(Q^+)^{\dagger}=Q^-$ and $(\bar Q^+)^{\dagger}=\bar Q ^-$.
Defining $Q_{\pm}=(1/\sqrt 2)(Q^{\pm} + \bar Q^{\pm})$ we have
\eqn\use{\{ Q_-,Q_+\} =H \qquad [Q_{\pm},H]=0.}

It is well known that on a non-compact space in general the central term
$\Delta$ can be non-zero \rWO. $\Delta$ depends on the boundary conditions at
spatial infinity: with multiple vacua, we have the freedom of having different
boundary conditions at left-right spatial infinity and thus different central
terms.  Let us denote the vacuum at left spatial infinity by $a $ and the one
on the right by $b $, so that the central term in the above algebra may be
labeled by $\Delta_{a b }$.  Having multiple vacua allows kinks which
interpolate from one vacuum at left spatial infinity to another one at right
spatial infinity.  The kinks we will denote by $k_{a b}$.  In general, such
kinks (even the lowest-energy configuration in each sector) may be stable or
unstable.  We can, however, derive a lower bound on the mass of any kink. In
the $ab$ sector the positivity of $\{ A, A^\dagger \}$ where $A= (H_R\bar
\Delta )^{1/2}Q^+-(H_L \Delta)^{1/2}\bar Q^-$ implies the Bogomolnyi bound
$E^2-P^2 \geq |\Delta_{ab} |^2$.  A kink $k_{ab}$ or, more generally, any
state in the $ab$ sector must therefore have mass $m\ge |\Delta_{ab}|$.

We now ask whether Witten's index is a good index for the $N$=2 case on open
space.  Consider varying the Hamiltonian of the theory, respecting $N$=2
supersymmetry.  We wish to compute
$$\delta {\rm Tr} (-1)^F e^{-\beta \rm H} =-\beta
{\rm Tr}(-1)^F \delta {\rm H}
e^{-\beta H}$$
Using \use\ we can write this as\foot{ We have been somewhat cavalier with
regard to boundary conditions at spatial infinity in taking the variations.
This point is elaborated upon in the appendix.}
$$-\beta{\rm Tr} (-1)^F \delta \{Q_+,Q_-\} e^{-\beta \rm H} $$
$$=-\beta {\rm Tr} (-1)^F (\{ \delta Q_+, Q_-\} +\{ Q_+, \delta
Q_-\})e^{-\beta H}$$
Each of the above terms vanishes.  To see this note that whenever we
are computing
$${\rm Tr}(-1)^F \{ A, B\}\  O$$
where $A$ and $B$ are fermionic and where at least one of them commutes with
$O$, we formally get zero. Suppose $A$ commutes with $O$.  Then for the $AB$
term contributing to the above trace, we can take $A$ around the trace,
because the trace is cyclic.  The term picks up a minus sign because $A$
anticommutes with $(-1)^F$.  This leaves $-BA$, which cancels $+BA$ from the
other term in the anti-commutator.  The same argument works if $A$ and $B$ are
bosonic operators, and we replace anti-commutators with commutators.  We shall
refer to this as the $AB$ argument.  For this formal argument to be actually
valid one needs to put restrictions on the nature of the operators $A$ and
$B$, which we assume to be satisfied in our case\foot{In the supersymmetric
quantum-mechanical version of this statement, this can be explicitly checked
to be true, where $A$ (after being dressed by $O$) is a trace-class operator
and $B$ (after being dressed by $O$) is bounded.} \mathr .  That the $AB$
argument is valid in our case is confirmed in the next section where we derive
the results of this section without making use of this assumption.

Applying these general statements to the above variation of the index, where
in one term $Q_-$ and in the other term $Q_+$ plays the role of $A$ in the
$AB$ argument, we find that the variation of Witten's index is zero for $N$=2
theories, and it is thus a good index even for non-compact space.  This in
particular means that in sectors where the left and right vacua are not the
same $\Tr(-1)^F e^{-\beta H}$ vanishes: we are free to take $\beta$ large
because it is an index, and since the ground state in this sector has non-zero
energy (because it interpolates between two distinct vacua) we get zero.

For an $N$=1 supersymmetric theory in two dimensions the fermion number $F$ is
only defined mod $2$.  However, in a two-dimensional $N$=2 theory there is a
$U(1)$ fermion-number charge, because the fermions are complex.  Given the
power of Witten's index in understanding the structure of supersymmetric
quantum field theories, it is thus natural to ask what kinds of objects may be
of interest when we have this additional charge.  The most natural thing to
consider would be
\eqn\parti{Z(\alpha ,\beta )={\rm Tr}\ e^{i\alpha F} e^{-\beta H}.}
At $\alpha =\pi$ this is just Witten's index.  For $\alpha =0$ it is just the
standard partition function of the theory, so we expect $Z(0,\beta )$ to be
the extreme opposite to an `index', as it should depend on {\it every little
detail} of the theory.  So let us go back to the point $\alpha =\pi$ and just
move {\it slightly away}.  In other words, consider
$$I_l(\beta )= {\partial ^l Z(\alpha ,\beta )\over \partial (i\alpha
)^l}\big|_{\alpha=\pi}= {\rm Tr}(-1)^F F^l e^{-\beta H}.$$
Needless to say, we should not expect all $I_l$ to be indices as that would
enable us to reconstruct $Z$ itself.  But maybe some of them are!  In
particular, consider $I_1={\rm Tr}(-1)^F Fe^{-\beta H}$.  Among all $I_l$ with
$l\geq 1$, we will show that this and only this is a new `index'.

To define what we mean by `index', we must recall that there are two
distinct ways to perturb an $N$=2 supersymmetric theory in two dimensions
\susy : $D-$terms and $F-$terms. In general, the $D-$terms can be written as
integrations of superfields over the full superspace $d^4\theta$ and the
$F-$terms which are integration of chiral and anti-chiral fields over half the
superspace $d^2 \theta^+$ and $d^2\theta^-$ respectively.  Chiral fields
commute with $Q^+$ and $\bar Q^+$ and anti-chiral fields commute with $Q^-$
and $\bar Q^-$.  The \ind\ is {\it independent} of the $D$-terms and in this
sense it is an index.  It however, {\it does depend} on the $F-$terms.  In
order to explain why we use the word `index' when it does depend on $F$-terms
it is convenient to consider the following interesting class of examples of
$N$=2 supersymmetric QFT's.  Consider 2d supersymmetric sigma models with
target space being a Kahler manifold $M$.  This gives an $N$=2 supersymmetric
theory \AlFr .  Any variation of the metric of $M$ respecting the complex
structure of $M$ and the Kahler class of the metric (i.e., leaving the
integral of the Kahler form on the two-cycles unchanged) can be written as a
$D-$term and so does not affect our index.  In fact this includes essentially
{\it all} possible perturbations of the manifold, modulo variations of complex
structure and Kahler structure which usually form a finite dimensional space
of perturbations.  So it is with this kind of example in mind that we call the
above object an `index'.  Another interesting class of $N$=2 supersymmetric
theories is provided by Landau-Ginzburg theories.  In these cases the
superpotential $W$ is the $F$-term and it has only a finite number of
perturbations which do not change the behavior of potential at infinity in
field space.  These turn out to be the relevant (and marginal) perturbations.
The index depends only on $W$.

 Here we show that \F\ does not depend on the $D-$terms.  The variation of the
$D-$term can be written as inserting $\{ Q^+, [\bar Q^-,\Lambda (x) ]\} $ in
the path integral where $\Lambda$ itself can be written as $\{ Q^-, [\bar Q^+
, K]\} $.  This follows from the fact that the $D-$term comes from integration
over all four Grassman coordinates.  The path-integral is over an infinite
cylinder of perimeter $\beta$ with the above term inserted at all points $x$
and integrated over the cylinder.  Let us denote by $\Lambda $ the integral of
$\Lambda (x)$ over space.  Since $F$ commutes with both $\Lambda $ and the
Hamiltonian we find that the integration of $\Lambda$ over the perimeter
simply introduces an irrelevant factor of $\beta$ which can be ignored.  So we
can write the variation of our index as (proportional to)
$$\delta ={\rm Tr}(-1)^F \ F\{ Q^+, [\bar Q^-,\Lambda ]\} e^{-\beta
\rm H}$$
We are almost ready to apply the $AB$ argument, using $Q^+$ as our $A$.  This
works fine, except for the fact that as we try to take $Q^+$ around the
trace,
since it does not commute with the $F$, we pick up a commutator term
$$\delta ={\rm Tr}(-1)^F [F,Q^+][\bar Q^-,\Lambda ] e^{-\beta \rm H}=
{\rm Tr}(-1)^F Q^+[\bar Q^-,\Lambda ] e^{-\beta \rm H}$$
Now we can apply an argument similar to the $AB$ argument, by taking $\bar
Q^-$ in the term $\bar Q^-\Lambda$ around the trace.  Here we pick up two
minus signs, and so we get back $\Lambda \bar Q^-$ which thus cancels the
second term in the commutator and we get zero, as was to be shown.

The $AB$ argument does not allow us to show that any of the other $I_l$ are
independent of $D$-term perturbations\foot{One can show using the $AB$
argument that the $D$-terms will not affect the one-particle contribution of
Bogomolnyi-saturated states to $I_l$, in accordance with the fact that the
mass and fermion number of these states are independent of $D$-terms.}.  In
fact, in a free massive $N=2$ theory, it is easy to compute $I_l$; the
$I_{2l}$ are non-vanishing and for $l\not= 0$ they do depend on the mass of
the particle, which in turn depends on the $D$-term.  In section 6 we consider
other examples for which all $I_{l}$ are non-vanishing and all of them for
$l>1$ depend on the $D$-term. Therefore $I_1$ is the only additional index
that exists other than Witten's index.  {}From now on we will refer to the new
index simply as $I$, dropping the subscript $1$.

Our index is actually a matrix, because we have to fix the boundary condition
at spatial infinities to be vacua of the theory.  If we choose the left vacuum
to be $ a$ and the right one to be $b$, we have the index $I$
as a matrix
$$I_{ab}={\rm Tr}_{a b}(-1)^F F\ e^{-\beta H}.$$
One has to be careful about what we mean by $(-1)^F$.  In general all that is
required from this operator is that it anti-commute with fermionic fields.  In
our case, as we have mentioned before, since $F$ is in fact well defined as an
operator, one can just define $(-1)^F=e^{i \pi F}$.  Note, however this
operator no longer squares to one.  The reason for this is that in the $(ab)$
sector the vacuum will in general have a non-integral fermion number $f_{ab}$.
This phenomenon is well known \refs{\rJR ,\rGW ,\rNS}. Only the fermion number
relative to that of the vacuum is integral.  Using this fact and the
hermiticity of $H$ and $F$ we can write
\eqn\pse{I_{ab}=\pm e^{i\pi f_{ab}}|I_{ab}|}

$CPT$ invariance puts constraints on our index.  $CPT$ takes a state in the
$(ab)$ sector to one in $(ba)$ sector, and it takes fermion number $F$ to
$-F$. In particular, $CPT$ invariance requires $f_{ab}=-f_{ba}$, and therefore
\eqn\cpt{I_{ab}=-I_{ba}^*}
There is no fractional fermion number in $(aa)$
sector: the fermion number is additive,  i.e.,
$f_{ac}=f_{ab}+f_{bc}$,
implying that $f_{aa}=f_{ab}+f_{ba}=0$. We see from \pse\ that $I_{aa}$ is
real, and from \cpt\ it follows that
\eqn\diag{I_{aa}=0.}
Note that if we had defined $(-1)^F=e^{(2n+1)i\pi F}$ then the index would
have changed by a phase $I_{ab}\rightarrow e^{2i n \pi f_{ab}}I_{ab}$.
However, under this ambiguity, the {\it eigenvalues} of $I$ are unambiguous.
Because of the additivity of the fractional part of the fermion number, and
since $f_{ab}=-f_{ba}$, we can write $f_{ab}=f_a-f_b$; a change of basis
$b\rightarrow e^{2i n \pi f_b}b$ gets rid of the phase without changing the
eigenvalues.

In fact, we can do better; we can get rid of all the phases (modulo $\pm$ ) of
our matrix by changing the basis $b\rightarrow e^{i \pi f_{b}}b$.  In this way
we find that $I$ is a purely real matrix, and the condition \cpt\ implies that
it is anti-symmetric. Thus its eigenvalues are either zero, or (purely
imaginary) complex-conjugate pairs.  To make the eigenvalues real, we define
the $Q$-index to be
\eqn\qmat{Q_{ab}={i\beta \over L}{\rm Tr}_{ab}(-1)^F F\ e^{-\beta H},}
where $L$ is the volume of the space. With this definition, $Q$ is a hermitian
matrix with real eigenvalues, such that non-zero eigenvalues come in pairs
of opposite sign.  To see the reason we divided by $L$ in the
definition of $Q$, consider $Z(\alpha ,\beta )$ (from \parti ) with boundary
conditions at infinity corresponding to a normalized eigenstate of $Q_{ab}$.
Because it is an extensive thermodynamic quantity, $\ln Z(\alpha ,\beta) =\ln
Tr (e^{i\alpha F}e^{-\beta H})$ is proportional to $L$ as
$L\rightarrow\infty$.  Therefore
\eqn\indd{{\beta\over L}\partial _{\alpha}\log \Tr (e^{i\alpha
F}e^{-\beta H})\big|
_{\alpha =\pi}={{i\beta\over L}\Tr F(-1)^Fe^{-\beta H}\over
\Tr(-1)^Fe^{-\beta H}}=Q ,}
where the denominator is not proportional to $L$, because it is Witten's
index, and can be chosen to be 1 in an orthonormal basis of eigenstates of
$Q$.  Thus we see that $Q$ as defined above is well-defined as
$L\rightarrow\infty$.

Usually the contribution of $n$ kinks (particles) to a partition function is
proportional to $L^n$.  One may incorrectly conclude from this that only the
one-kink states contribute to $Q$.  In fact we will see in later sections that
the $n$-kink contributions to $I$ do not generally vanish and are proportional
to $L$.  The contribution comes from regions where all the kinks are near each
other and the factor of $L$ is associated with the center of mass.  It can be
seen that any configuration where one of the kinks is very far from the rest
does not contribute to $Q$: the contribution factorizes and at least one piece
will simply be the contribution to Witten's index from massive kinks, which
vanishes.  One can also see this from the path-integral computation where the
exact fermion zero modes associated to each kink when they are
far away cannot be absorbed by one $F$.

In the next section we will see that $Q$ is the same as the matrix element of
the chiral fermion number:
$$Q_{ab}=\langle a|Q^5|b\rangle .$$
Using this expression along with the hermiticity of $Q^5$, and noting
that $CPT$ changes the sign of $Q^5$, we again see that the eigenvalues
of $Q$ are real and symmetrically located relative to zero.

For the remainder of this section, we will discuss the kind of states in the
Hilbert space which contribute to our index.  In general, there are three
types of irreducible representations of the supersymmetry algebra \sual .  The
generic irreducible representation of \sual\ is four-dimensional, with a
definite eigenvalue for $E$ and $P$ (as $H_{L,R}$ commute with everything).
This follows from the fact that the four supersymmetry charges which generate
the algebra are pairwise adjoint of one another and have $c$-number
anti-commutators.  We can generate the representation by taking $Q^+$ and
$\bar Q^-$ as `creation' operators acting on a state which is annihilated by
the `annihilation' operators $Q^-$ and $\bar Q^+$:
\eqn\nred{|s\rangle \quad Q^+|s\rangle \quad \bar Q^- |s\rangle \quad
Q^+\bar Q^- |s\rangle }
When $E^2-P^2=\Delta \bar \Delta$, i.e., if the state saturates the Bogomolnyi
bound, then it is well known \rWO\ that this representation is reducible:
$A=(H_R\bar \Delta )^{1/2}Q^+ - (H_L\Delta )^{1/2})\bar Q^-$ and its adjoint
anticommute, and so both must annihilate $|s\rangle $. This leaves us with the
reduced supersymmetry multiplet
\eqn\red{|s\rangle \qquad Q^+|s\rangle }
Finally, for $E=P=\Delta=0$ this representation is further reduced to
the trivial representation.  This representation only appears for the $(aa)$
sectors, and are the only states which contribute to
Witten's index in this sector.  However, because $I_{aa}=0$
these states are not relevant for the new index $I$.

At first glance, one might think that only the reduced multiplets contribute
to our index.  If $|s\rangle $ has fermion number $f$, a non-reduced multiplet
\nred\ naively contributes (up to an overall phase)
 $(f-2(f+1)+(f+2))e^{-\beta E}=0$, whereas a reduced multiplet \red\
contributes (up to an overall phase) $(f-(f+1))e^{-\beta E}=-e^{ -\beta E}$.
Thus it appears that $I$ receives contributions only from Bogomolnyi-saturated
states, which are simply the one-soliton subsectors.  This argument is
incorrect, for the same reason that the naive argument which states that ${\rm
Tr}(-1)^F e^{-\beta H}$ is independent of $\beta$ is not in general valid when
$H$ has a continuous spectrum, as is the case in non-compact spaces.
Formally, we have deduced the vanishing of the contribution of the non-reduced
multiplets only when the spectrum of the Hamiltonian is discrete.  When it is
continuous, as with a model on a real line, we have to deal with the {\it
density} of states of the non-reduced multiplets; they are not necessarily
equal and do not necessarily cancel in computing $I$.  We may wish to
regularize the theory by putting it in a box of size $L$ and then take
$L\rightarrow \infty$.  In order to recover the soliton sector $ab$, the field
configurations on the left and the right of the box (in this case just a line
interval) cannot be the same.  Thus we {\it cannot} impose periodic boundary
conditions.  We must compute the object in finite but not periodic box, and
this breaks the supersymmetry. The spectrum is discrete in this case, but
without supersymmetry the naive argument no longer holds.  Thus for a finite
box we may get contribution from non-reduced supersymmetry multiplets to the
index $Q$ in the sector $ab$ with $a\not= b$.  This may persist even when the
size $L\rightarrow \infty$ \foot{ This is indeed one way that the
$\beta$-dependence of ${\rm Tr}(-1)^F e^{-\beta H}$ has been computed in the
supersymmetric quantum mechanics examples \sqm , which is related to the
Callias-Bott-Seeley index.}.  Thus we are computing a kind of `anomaly', which
remains after the regulator is removed.

Let $g_f(E)$ be the density of states for $|s\rangle$, $2g_{f+1}(E)$ be the
density of states for states spanned by $Q^+|s\rangle$ and $\bar
Q^-|s\rangle$, and $g_{f+2}(E)$ be the density of states for states $Q^+\bar
Q^-|s\rangle$. We should thus not expect the continuum densities $g_f(E)$,
$g_{f+1}(E)$, and $g_{f+2}(E)$ to be equal in the $(a,b)$ sector of the theory
with $a\neq b$.  Recall, though, that we proved using $N$=2 supersymmetry that
the contribution to Witten's index from these states must cancel.  This means
that we must have
\eqn\ivreln{g_f(E)+g_{f+2}(E)=2g_{f+1}(E),}
since the states on the two sides make opposite contributions to
Witten's index.  Using \ivreln\ we see that the contribution of the
four dimensional representation to the index $I$ in the $(a,b)$
sector is of the form
\eqn\ivdiml{e^{i\pi f}\int dE(g_{f+2}(E)-g_{f}(E))e^{-\beta E}.}
  We will see explicitly how this is generically nonzero in the following
sections.

\newsec{Geometry of Ground States and the New Supersymmetric Index}
In this section we review some aspects of the work done in \cv\ which are
useful for the considerations of this paper.  In particular, we show why the
`Q'-matrix discussed there is in fact the new supersymmetric index given by
$\Tr (-1)^F F e^{-\beta H}$ discussed in the previous section.  It is
convenient to exchange the role of space and time (i.e., do a `modular
transformation') and take the space to be a circle (with perimeter $\beta$ to
correspond to the index computation), with periodic boundary conditions.  Time
is now a line of length $L$.

Consider an arbitrary $N$=2 supersymmetric quantum field theory in two
dimensions.  {}From \use\ and the positivity of the inner product together with
the fact that $Q_-=Q_+^\dagger$, it is easy to show that the ground states of
the theory are characterized by
$$H|a\rangle =0 \leftrightarrow Q_{\pm}|a\rangle =0.$$
There is thus a one-to-one correspondence
between the ground states of the theory and the $Q_+$ or $Q_-$ cohomology.
This cohomology is definable because each of these operators squares to zero
(note that on a compact space (circle) the supersymmetry algebra has no
central term and we get $Q_+^2=Q_-^2=0$).  The analogy to keep in mind is that
$Q_+$ is like a $d$ operator acting on the differential forms on a manifold,
$Q_-$ is like the adjoint operator $d^\dagger$ and the ground states $|a
\rangle $ are like the harmonic representative of $d$ or $d^\dagger$
cohomology.

In correspondence with the ground states in the Hilbert space, there are
chiral operators $\phi_i$ in the theory defined by the condition that
$$[Q_+,\phi_i ]=0$$
and similarly there are anti-chiral operators $\bar \phi_i$ which commute with
$Q_-$.  Acting on a vacuum by a chiral operator, we get another state which is
$Q_+$ closed, another $Q_+$ cohomology element.  In this
way the chiral fields, modulo the fields that are trivially chiral, i.e.
modulo fields which are themselves $Q_+$ (anti-)commutator, are in one-to-one
correspondence with the $Q_+$ cohomology elements and thus the ground states.
If we pick a canonical ground state (to be defined below) denoted by
$|0\rangle$, this can be stated as
$$\phi_i |0\rangle =|i\rangle +Q_+|\Lambda \rangle$$
where $|i\rangle $ denotes another ground state.  Similarly we can label the
ground states using the anti-chiral fields $\bar \phi_i$ which leads to the
states $|\bar i \rangle$.  The chiral fields form a ring among themselves,
called the chiral ring, which is defined by
$$\phi _i \phi_j =C_{ij}^k \phi_k +[Q^+, \Lambda ]$$
$$\phi_i |j\rangle =C_{ij}^k |k\rangle .$$
The matrix $(C_i)_j^k=C_{ij}^k$ denotes the action of the chiral field
$\phi_i$ on the ground states (once we ignore the components orthogonal to
ground states).  Similar statements apply to anti-chiral fields with
$C_{ij}^k$ replaced by the complex conjugate quantity $(C_{ij}^k)^*$.

We can define a symmetric metric $\eta$ and a hermitian metric $g$ among the
ground states by
$$\eta_{ij}=\langle i|j\rangle \qquad g_{i\bar j}=\langle \bar j |i\rangle $$
Note that the metric $g$ is the usual metric in the Hilbert space of the $N$=2
theory and $\eta$, which is not hermitian, is a kind of `topological' metric.
As discussed in the previous section there are two ways to perturb the action:
the `D-terms' (denoted by $K(X,\bar X)$ below) and the `F-terms' which are the
chiral fields (now viewed as superfields) and integrated over half of the
superspace:
$$S\rightarrow S+\int d^2z d^4 \theta K(X,\bar X) +
\int d^2z d^2\theta ^+ \ t_i \phi_i + \int d^2z d^2\theta^-
\ {\bar t_i}{\bar \phi_i}.$$
Then it is possible to show (see \cv ) that the chiral ring and the metrics
$\eta $ and $g$ {\it depend only on the} F-{\it terms}, i.e. they depend only
on $t_i,\bar t_i$ and are independent of $K$.  The flavor of the argument is
very similar to the argument in the previous section in showing that our index
is independent of $D-$terms, but it has the advantage of being
rigorous.

The ring matrices $C_i$ and the metric $\eta$ can also be related
to computations of correlation functions in a topological theory \witten\
corresponding to `twisting' the $N$=2 quantum field theory and can
thus be easily computed exactly \refs{\witten, \eg , \dvv , \tlg}.
 Basically the topological theory is the same as the ordinary $N$=2
theory on flat manifolds but differs from it when the two-dimensional manifold
is not flat, in such a way that the charge $Q_+$ transforms as a scalar, and
is thus a symmetry even if the space is not flat.  The way this is
accomplished is by introducing a background gauge field
set equal to half the spin connection of the manifold, and coupling
it to the fermion number current. Thus a
field which previously had spin $s$ and fermion
charge $q$ will now have spin $s-{1\over 2}q$.  This in particular makes $Q_+$
which had spin $1/2$ and fermion number $+1$, a scalar.  If $S$ denotes the
action of ordinary $N$=2 theory, $S_t$ denotes the action for the
topological theory, $j$ denotes the fermion number current, and
$\omega_\mu$ denotes the $U(1)$ spin connection we have
\eqn\topa{S_t=S+{i\over 2} \int j_{\mu} \omega^\mu .}

An important property of the topological action is that the energy-momentum of
the topological theory is itself $Q_+$ trivial:
\eqn\ener{T^t_{\mu \nu}=T_{\mu \nu}+{1\over 2}
\epsilon_{\alpha (\mu}\partial_{\nu )} j^\alpha =\{ Q_+ ,\Lambda \} }
implying that the correlation functions for chiral fields are independent of
the metric.  By translating the computation of the $N$=2 topological theory
into the language of the ordinary $N$=2 theory, this provides exactly the
quantities $\eta$ and the ring matrices $C$.  The basic observation is that if
we consider a hemisphere and do the path-integral in the topological theory we
get a state (on the boundary circle) which is annihilated by the symmetry
charge $Q_+$.  Moreover because the energy momentum tensor is $Q_+$-trivial,
any local variation of the data (such as the variation of the metric on the
hemisphere) does not change the $Q_+$ cohomology class of the state, and so
the path integral of the topological theory leads to a well-defined state in
the $Q_+$-cohomology, and thus to a ground state of the ordinary $N$=2 theory.
In particular the state that we called the vacuum $|0\rangle$ corresponds to
the state we get when we do the path-integral
with no insertion of any fields on
the hemisphere. Simple arguments show that $C$ and $\eta$ depend only on $t_i$
and not on $\bar t_i$.  In other words, they are holomorphic.

Similarly, we can consider the {\it anti-topological} theory, which is
obtained when we make $Q_-$ a scalar.  This is done simply by changing the
sign of the background field, which shifts the spins by $s\rightarrow
s+{1\over 2}q$.  So the action for the anti-topological theory $S_{t*}$ is
$$S_{t*}=S-{i\over 2} \int j_{\mu} \omega^\mu $$
{}From the anti-topological theory we can easily compute $\eta_{\bar i\bar j}$
and $\bar C$ which are simply the complex conjugate of the corresponding
topological quantities $\eta $ and $C$.

The computation of the hermitian ground-state metric $g$ as a function of
perturbation parameters $(t_i,\bar t_i)$ is more difficult.  It turns out that
by fusing the topological theory on one hemisphere with the anti-topological
theory on the other hemisphere, we can find equations which characterize it
\cv .  This we shall call topological-anti-topological fusion, or $tt^*$ for
brevity.  One simply introduces a gauge connection such that the variation of
ground states are orthogonal to the ground states themselves:
$$D_i|a \rangle =\partial_i -A_i |a \rangle
\qquad \bar D_i |a \rangle =\bar \partial_i -\bar A_i |a \rangle .$$
This in particular means that the metric $g$ is covariantly constant
$$D_i g=\bar D_i g=0,$$
and one finds the equations
$$[D_i,D_j]=[\bar D_i,\bar D_j]=0$$
\eqn\curv{[D_i,\bar D_j]=-\beta^2 [C_i, \bar C_j]}
(and some other equations which we will not need here).  The perimeter of the
space (circle) is $\beta$.  We will give a quick (but not rigorous) derivation
of the above equations in the spirit of the $AB$ argument of previous section
in the appendix.

 The first equation \curv\ shows that we can choose a holomorphic gauge with
$\bar A_i=0$.  This turns out to be the natural gauge in the topological
theory.  In more mathematical terminology we can say that the topological
path-integral automatically gives holomorphic sections of the vacuum bundle.
Using the covariant constancy of the metric we can write the metric $g$ as
$$A_i=-g \partial _i g^{-1}$$
and so the second equation in \curv\ becomes
$$\bar \partial_j (g \partial_i g^{-1})=\beta^2[C_i, g C_j^\dagger g^{-1}]$$
In many examples these equations turn out to be among the celebrated equations
of mathematical physics.  For the $N$=2 sine-Gordon theory the above equation
as a function of the scale turns out to correspond to radial solutions of the
sinh-Gordon differential equation, which is a special case of Painleve III.
These differential equations are always integrable, being related to a tau
function.  The integrability of these equations has been recently elaborated
upon in \dub .  Explicit numerical computations have been done for
flows among conformal theories and also flows under generic perturbations away
from conformal theories \bill .

Among the perturbations of the $N$=2 theory, there is a special one
corresponding to renormalization group flow.  In particular if we denote the
perimeter of the circle on which we base our Hilbert space as $e^\tau$, then
changing $\tau $ should be equivalent to changing the coupling in the theory
in some particular way.  In the case of Landau-Ginzburg theories, this has the
same effect on the F-terms as multiplying it by $e^\tau$.  {}From the
definition of connection it follows that
$$\partial_\tau |a \rangle =A_\tau |a \rangle $$
On the other hand it was shown in \cv\ that the variation of the ground states
with respect to the perimeter is related to the action of the chiral fermion
number charge $Q^5$ on the ground states by
\eqn\chir{\partial_\tau |a \rangle ={1\over 2} (Q^5+ n)|a \rangle }
where the above equality holds as long as we project both sides back to the
ground states.  Here $n$ is a number which measures the chiral anomaly of the
theory (equal to the number of chiral fields in the LG theory). So we see that
as far as the ground state action is concerned, in a holomorphic (topological)
basis\foot{More precisely, in a topological basis $|i\rangle$ obtained by
inserting in the topological integral chiral operators $\phi_i$ with
$\partial_\tau\phi_i=0$. Two such bases are related by a $\tau$ independent
`gauge transformation'. Under such changes of bases $g\partial_\tau g^{-1}$
transforms as a tensor.}
\eqn\main{{1\over 2}(Q^5+n)|i\rangle =A_{\tau\ i}^{\ j}|j\rangle =
(-g \partial_\tau g^{-1})_i^{\ j}|j\rangle }
The equation \chir\ was derived in \cv\ in the context of Landau-Ginzburg
theories.  Since this is an important equation for us in this paper, we will
now present a more general derivation of it.

It is convenient to work in the topological basis.  Then a state $|i \rangle$
can be obtained as a result of topological path-integral on hemisphere, with
insertion of the chiral field $\phi_i$.
In view of the fact that the energy momentum
tensor of the topological theory is $Q_+$ trivial, it sounds contradictory to
expect $\partial_\tau |i \rangle $ not to be zero (i.e., $Q_+$ trivial).  The
way this comes about is by a subtle boundary term, as we will now see.

Let us denote the metric on the hemisphere by $h=e^{2\phi} dz d\bar z$.
In terms of $\phi$ the spin connection is $\omega =* d\phi $ and so the
topological action \topa\ is
$$S_t=S+{i\over 2}\int j \wedge d\phi .$$
Now we are interested in the variation of this action on the right hemisphere
as we change $\phi $ by a constant.  Varying the metric by an overall scale
$\phi \rightarrow \phi +\epsilon$ has
the effect of changing the perimeter by shifting $\tau\rightarrow\tau
+\epsilon $.  It is convenient to first do a partial
integration on the second term above and write it as
$$\int_{S_R} j\wedge d\phi =\oint_{S^1}j\phi -\int_{S_R}\phi dj$$
where $S_R$ denotes the right hemisphere and $S^1$, the boundary circle of
$S_R$, is where we base our Hilbert space.  Shifting $\phi$ brings down from
$S$ the trace of the energy momentum tensor, and from the topological addition
the divergence of the axial current plus the variation of the boundary term,
i.e.,
$$\delta S=\int_{S_R }(T_\mu ^\mu + {1\over 2} D_\mu j^{5 \mu}) +{i\over 2}
\oint _{S^1} j$$
where we have used that in {\it two}-dimensional Euclidean field theory
$j^5_\mu =i\epsilon_{\mu \nu} j^\nu$. The term integrated over the right
hemisphere appears to be $Q_+$ trivial because it is the trace of the energy
momentum tensor of the topological theory.  This statement is almost true,
except for the fact that there is a well-known anomaly in the divergence of
axial current which contributes $n/2$ (in the LG theory $n$ is the number of
fields).  But now we see that the boundary term is also present, and is
equivalent to the action of $Q^5/2$ at the boundary (as follows from $j^5=i
(*j)$).  So the net effect on
a state of the change of $\tau$ is given by
$$\partial_\tau |a \rangle = {1\over 2} (Q^5 +n)|a \rangle $$
(as long as we compute the matrix element of both sides of
the above equation among ground states).  This is the equation \chir\ we
wished to derive.

We have seen that the matrix elements of $Q^5$ among ground states of the
supersymmetric theory are possible to compute, if we know $g$ (from \main ).
Note that even though the fermion number is always conserved the chiral
fermion number is conserved only at conformal points.

Since we are considering both massive and massless theories it may seem
strange to see that the matrix elements of a non-conserved charge are somehow
`interesting' and related to RG-variations of ground states.  Let us rephrase
this by using a modular transformation.  Consider the theory on a very long
cylinder of length $L$ and circumference $\beta $.  Let us put a ground state
$| b \rangle $ at one end of the cylinder and another ground state $\langle a
|$ at the other end.  We denote the coordinates along the cylinder by $x$ and
that along the circumference by $t$.  The matrix elements of $Q^5$ can then be
written as
$$\langle a |Q^5|b\rangle =\langle
a |i\oint_{S^1} j_t(0,t) dt |b \rangle $$
where $S^1$ is a circle wrapped around the middle of the long cylinder.  We
have to take the limit $L\rightarrow \infty$ at the end in order to project
onto the ground states in a natural way.  It clearly does not matter where we
insert the circle.  So let us put the circle at any $x$, integrate over all
$x$ and divide by $L$, i.e.,
\eqn\naz{\langle a |Q^5 |b \rangle ={i\over L}
\langle a |\int j_t(x,t) dxdt |b \rangle }
Now viewing $x$ as space, and $t$ as time, we see that $\int j_t(x,t) dx$ is
the definition of the fermion number $F$ on the Hilbert space which is along
the cylinder.  Since fermion number is conserved, integrating along $t$ will
just introduce an additional factor of the circumference of the cylinder
$\beta$.  In other words we have
$$\int j_t(x,t)dx dt =\beta F$$
So we have finally
\eqn\mod{Q_{a b}=\langle a  |Q^5 |b \rangle =
{i\beta \over L}{\rm Tr}_{a b}(-1)^F F\ e^{-\beta H}}
where the ${\rm Tr}_{a b} $ means that we are taking the boundary conditions
on the left and right to correspond to $\langle a |$ and $|b \rangle$ vacua.
This is the new index discussed in the previous
section! What is surprising is that the index
can be computed exactly in terms of $g$, and $g$ is determined
exactly by the differential equations \curv .  In particular we see from
\main\ that the index is given by
\eqn\final{Q_{ab}={i\beta \over L}\ {\rm Tr}_{a b}(-1)^F F\ e^{-\beta H}=
-(\beta g \partial_\beta g^{-1} +n)_{ a b },}
where we have used $2g\partial_\tau g^{-1}=\beta g\partial_\beta g^{-1}$ which
follows because, by scaling, we can set $\beta =e^{\tau /2+\tau ^*/2}$.  Often
it is difficult to compare the topological basis for ground states with the
path-integral choice emphasized in the previous section and more natural from
the viewpoint of kinks. In such cases it is convenient to compare the {\it
eigenvalues} of the $Q$ matrix on both sides of the above equation.

 Note that the matrix
$Q$, since it can be written solely in terms of $g$,
 depends only on the knowledge of the $F-$term and is independent of the
$D$-terms in accord with our proof in the previous section.
Our final formula, \final , expresses the new supersymmetric index in terms of
the geometry of supersymmetric ground states.  Because the curvature of this
space is determined simply from the chiral ring structure constants using
\curv , the index will be an exact solution of a differential equation whose
form is determined simply by the chiral ring.  In other words, though our
index is not purely topological, its flow in $\beta$ is determined using only
topological data, namely the chiral ring.

At the conformal point, where chiral fermion number is conserved, $Q$ measures
the chiral charge of Ramond vacua, i.e., the left-moving fermion number plus
the right moving fermion number.  In this case the state with highest charge
has $Q={\hat c}$ where $\hat c$ is the central charge of the $N$=2
superconformal theory \refs{\lvw ,\rVW }.
So off criticality each eigenvalue of the
$Q$ matrix, and in particular the highest
one, is a kind of a generalization of a
$c$-function \cpg (which has no
direct relationship with Zamolodchikov's definition \rZamc , as discussed
in \cv ).

\newsec{The infra-red expansion of \F}

In section 2 we discussed which states in the Hilbert space contribute to
\break \F .
In this section we show how to calculate the one- and two-particle
contributions. These are the leading terms in infra-red limit where $\beta
>>1$.  We will see the simple but non-trivial nature of our index. These
results must be the leading infra red behavior
of the $tt^*$ differential equations of the previous section.

Let us start with the contribution of one-particle (kink) states to the index.
In order to calculate the density of states, we put the system in a box of
length $L$ with the $a$ and $b$ boundary conditions at the end of the box.  To
obtain a non-vanishing contribution to the index, recalling \diag , we take
$a\neq b$; in particular, we do not want periodic boundary conditions.  The
allowed momenta of a particle in a box are quantized as $p=n\pi/L$, where $n$
is a positive integer. Thus the density of states for each component of a
supersymmetry multiplet is the same and given by $g(E)dE ={L dp/ \pi}$. {}From
relation \ivdiml , we see that one-particle states in four-dimensional
multiplets do not contribute to the index; a single particle contributes if
and only if it is part of a reduced supersymmetry multiplet.  This in
particular means that its mass should saturate the Bogomolnyi bound
$m_{ab}=|\Delta_{ab}|$.  So the one-particle contribution to $Q_{ab}={i\beta
\over L}{\rm Tr}_{ab} (-1)^F F e^{-\beta H}$ from a kink multiplet with
 fermion number $(f_{ab},f_{ab}+1)$ is given by
\eqn\ipart{\eqalign{&i \beta  (f_{ab}-(f_{ab} +1))
e^{i\pi f_{ab}}\int_0^{\infty}
{dp\over \pi}e^{-\beta \sqrt{p^2+m_{ab}^2}}\cr =&-i|\Delta_{ab}| \beta e^{i\pi
f_{ab}}{1\over\pi} K_1(|\Delta_{ab}|\beta),\cr}}
where $K_1$ is a Bessel function.  This simple statement explains and makes
precise the observation made in \cv\ that in the infra-red the Q-matrix is a
kind of partition function of the solitons of the theory.  The fact that the
leading term in the infra-red limit is proportional to $K_1$ follows easily
from the $tt^*$ equations (see appendix $B$ of \cv ).

The next-leading contribution in the infrared to the index comes from the
two-particle states.  A two-particle state generally forms one or more
four-dimensional non-reduced supersymmetry multiplets. This is true even if
both particles are individually reduced-multiplet, unless
$m_1+m_2=|\Delta_1+\Delta_2|$.  Thus the two-particle state generally{\it does
not} saturate the Bogomolnyi bound.  This is the first case where we can check
whether we get contributions of the form \ivdiml\ from four-dimensional
representations.  We will see that the two-particle contribution is very
simple and general, and often not zero, for the case where both particles are
part of reduced multiplets.

Computing the two-particle contribution is easy if one knows the two-particle
$S$-matrix; the $S$-matrix encodes the density of states \rLL. In a large box,
the particles spend a negligible amount of phase space near each other, so the
exact details of the interaction are unnecessary.  The $S$-matrix allows one
to match the free-particle solution of the  equation with
$x_1>>x_2$ with the one for $x_1<<x_2$.

Consider a two-particle state $|i(p_1,p_2)\>$ which scatters entirely into
another state $|j(p_1',p_2')\>$ with $S$-matrix element $S_{ij}(p_1,p_2)$.
\foot{We neglect processes which take two particles to more than two
(which should be a good assumption in the infrared limit).} Relativistic
invariance ensures that $S_{ij}$ actually depends only on $s\equiv
(E_1+E_2)^2-(p_1+p_2)^2=m_1^2+m_2^2+2m_1m_2
\cosh(\tht_1-\tht_2)$, where we define rapidities via $p=m\sinh\tht$.
Generically, $|i\>\ne |j\>$ for solitons even in elastic forward scattering,
where the individual $\Delta$ change. Consider a two-particle wavefunction
connecting vacua $a$ and $b$ at the box ends.  Properly-matched plane-wave
states satisfy
\eqn\planewave{ \psi(x_1,x_2)=\cases{e^{ip_1 x_1 + ip_2 x_2}&$
 x_1<x_2$,\cr
 e^{ip_1' x_1 + ip_2' x_2}S_{ij}(\tht_2-\tht_1)&$
  x_1>x_2.$\cr}}
The momenta can change in the collision; the relations $p_1+p_2=p_1'+p_2'$ and
$E_1+E_2=E_1'+E_2'$ give us the final momenta in terms of
the initial. Thus we can write $p_1'=p_1'(p_1,p_2)$.

Since our system is in a box of length $L$, an allowed wavefunction must
vanish at the walls. This quantizes the momenta just as in the free case,
but here the two quantization relations are coupled.  Requiring the
wavefunction vanish at $x_1=0$ means making a standing wave by subtracting the
solution with opposite $p_1$. When making it vanish at $x_1=L$, we use the
second relation in \planewave, and it follows that\foot{We also need to define
the states so that when $|i(p_1,p_2)\>$ scatters only into $|j(p_1',p_2')\>$,
then $|j(p_1'(-p_1,p_2),p_2'(-p_1,p_2))\>$
also scatters only into $|i(-p_1,p_2)\>$.  In other
words, the scattering remains diagonal even after a particle bounces off the
wall.}
\eqn\constr{e^{ip_1'(p_1,p_2) L}S_{ij}(\tht_2-\tht_1)=e^{ip_1'(-p_1,p_2) L}
S_{ij}(\tht_2+\tht_1),}
where we note that $p_1'(-p_1,p_2)$ is not necessarily equal to
$-p_1'(p_1,p_2)$.  Requiring the vanishing at $x_2=0$ and $x_2=L$ gives
another equation:
\eqn\constrii{e^{-ip_2'(p_1,p_2) L}S_{ij}(\tht_2-\tht_1)=e^{-ip_2'(p_1,-p_2) L}
S_{ij}(-\tht_2-\tht_1),}
Taking the log of \constr\ gives
\eqn\pconstr{2{n\pi}= k_1 L
+Im\ln{S_{ij}(\tht_2-\tht_1)\over S_{ij}(\tht_2+\tht_1)},}
where $n$ is an integer, and we define the kinematic factors
$$k_1=p_1'(p_1,p_2)-p_1'(-p_1,p_2) \qquad\qquad
k_2=p_2'(p_1,p_2)-p_2'(p_1,-p_2)$$
(notice that for forward elastic scattering, $k_i=2p_i$).
Taking the log of \constrii\ gives another relation:
\eqn\pconstrii{2{\bar n\pi}=k_2 L
-Im\ln{S_{ij}(\tht_2-\tht_1)\over S_{ij}(-\tht_2-\tht_1)},}

The contribution to the index from the two-particle state $i$ comes from
summing over all integers $n$ and $\bar n$, so that $p_1$ and $p_2$ are
greater than zero. Since we have free on-shell states, the energy is just the
free-particle energy.  The levels are close together because the box is large,
so we replace these sums with integrals. We also make the integral over
$\tht_1$ and $\tht_2$, so that we must multiply by the density of states
$g_{f_i}$, which is the Jacobian
\eqn\gfi{g_{f_i}={\partial n \over
\partial \tht_1}{\partial \bar n \over \partial \tht_2} -{\partial n \over
\partial \tht_2}{\partial \bar n \over \partial \tht_1}.}
The relations \pconstr\ and \pconstrii\ give ${\partial n /\partial \tht_i}$
and ${\partial \bar n /\partial \tht_i}$ each as the sum of two terms, one
proportional to $L$ (the ``free'' piece)
and the other involving the $S$-matrix.
Thus $g_{f_i}$ has a piece proportional to $L^2$. This results in the
two-free-particle contribution to the index. However, summing over each
four-dimensional representation gives a contribution of the form \ivdiml, and
the $L^2$ piece vanishes in $g_{f+2}-g_f$. We know this must happen, from the
discussion following \indd.  The contribution proportional to $L$ from a state
with fermion number $f_i$ is
\eqn\Qiisimp{\eqalign{f_i e^{i\pi f_i}
{L\over 2(2\pi)^2}\int\int d\tht_1 d\tht_2
&\left(({\partial\over\partial\tht_1}+
{\partial\over\partial\tht_2})(k_1+k_2)\right)\cr
&{\partial\over\partial\tht_1} Im \ln S_{ij}(\tht_2-\tht_1)
e^{-\beta (m_1\cosh\tht_1+m_2\cosh\tht_2)}.\cr}}
To simplify the expression, we have rewritten the rapidity integrals to
go from $-\infty$ to $\infty$ by using the fact that
$S(-\tht)=S^*(\tht)=1/S(\tht)$, which follows from analyticity and unitarity
of the $S$-matrix.

We now specialize to the case where both particles are in reduced
multiplets\foot{It is
perfectly conceivable that only configurations comprised of
particles belonging to reduced
multiplets contribute to the
index.}.
The result simplifies remarkably, and only depends on the individual
$\Delta$'s of the particles and not on any
details of the $S$-matrix.  We decompose
the initial states into four-dimensional supersymmetry
representations $i$ with fermion numbers ($f_i,f_i+1,f_i+1,f_i+2$) as in
\nred.  Denoting a reduced multiplet by ($d,u$), a two-particle state with
fermion number $f_i+2$ is of the form $|u_1 u_2\>_i$, while the one with
fermion number $f_i$ is $|d_1 d_2\>_i$.  By fermion-number conservation and
supersymmetry, $|u_1 u_2\>_i$ must scatter only into a state $|u_1'u_2'\>_j$
and likewise $|d_1 d_2\>_i$ must scatter only into a state $|d_1' d_2'\>_j$.
We denote the corresponding $S$-matrix elements by $a_{ij}$ and $\tilde a_{ij}$
respectively.

The relation \ivdiml\ means that we do not need to calculate all of the
densities: we only need the difference $g_{f_i+2}-g_{f_i}$.  Looking at
\Qiisimp, we see that the index thus depends only on the ratio
of $S$-matrix elements $a_{ij}/\tilde a_{ij}$. The striking fact is that this
ratio can be found without knowing the full $S$-matrix; it follows from the
supersymmetry alone.  We know that $Q^+\overline Q^-|f_i\>= \l_i |(f+2)_i\>$,
where $\l_i$ depends on the details of the representation $i$. The $S$-matrix
commutes with the supersymmetry generators, which means that the diagram
\def\joinrel{\mathrel{\mkern-3mu}}
\def\relbar{\mathrel{\smash-}}
\def\llong{\relbar\joinrel\relbar\joinrel\relbar\joinrel\rightarrow}
$$\eqalign{|d_1& d_2\rangle \hskip.2in
{\buildrel \tilde a \over \llong}\hskip.2in
|d_1' d_2'\rangle\cr
Q^+\overline Q^-&\Biggl\downarrow
\hskip 1in \Biggl\downarrow Q^+\overline Q^-\cr
|u_1 &u_2\rangle \hskip.2in{\buildrel  a \over \llong }
\hskip.2in|u_1' u_2'\rangle \cr }$$
must commute. This implies that
\eqn\aatil{{a_{ij}\over \tilde a_{ij}}={\l_j\over\l_i}.}

We can find the $\l_i$ from one-particle information.
The supersymmetry is represented on a doublet with $m=|\Delta|$ as
\eqn\SUSYrep{\eqalign{Q^-|u(\tht)\rangle &=\sqrt{m}e^{\tht/2}|
d(\tht)\rangle\qquad\qquad \bar Q^+|u(\tht)\rangle=\omega
\sqrt{m}e^{-\tht/2}| d(\tht)\rangle\cr Q^+|
d(\tht)\rangle&=\sqrt{m}e^{\tht/2}|u(\tht)\rangle\qquad\qquad \bar Q^-|
d(\tht)\rangle=\omega^*\sqrt{m}e^{-\tht/2}| u(\tht)\rangle,\cr}}
where $\omega =\Delta/|\Delta|$.  All other actions annihilate the states.
The supersymmetry is defined on multi-particle states in the usual manner.
Since $Q$ is fermionic, one picks up phases when $Q$ is brought through a
particle with fermion number. For example, bringing $Q$ through a fermion
results in a minus sign. Since we have fractional charges, we must generalize
this notion, so that the action of $Q^{\pm}$ on the tensor product of two
states is
\eqn\QQG{Q^\pm \otimes 1 + e^{\pm i \pi F}\otimes Q^{\pm}.}
The charges $\bar Q^\mp$ act with
the same phases as $Q^\pm$.  In our two-particle case of interest, we have
\eqn\pqpm{\eqalign{Q^+\overline Q^-&|d_1(\tht_1)d_2(\tht_2)\>\cr
&=e^{i\pi f_1}\sqrt{m_1m_2}
\left( \omega_1^* e^{(\tht_2-\tht_1)/2}-
\omega_2^* e^{(\tht_1-\tht_2)/2}\right)     |u_1(\tht_1)u_2(\tht_2)\>.\cr}}
The quantity of relevance in \Qiisimp\ is thus
\eqn\aatilsat{{\partial\over \partial\t _1}\ln{a_{ij}\over\tilde
a_{ij}}={\partial\over \partial\t _1}\ln
{\sinh({{\tht_2-\tht_1}\over 2}+\mu)\over \sinh({{\tht_2'-\tht_1'}\over
2}+\mu')},}
where $\mu=\half \ln \omega_2 \omega_1^*$. The contribution of two reduced
multiplets to the index $Q$ thus depends only on $\Delta_1$, $\Delta_2$,
$\Delta_1'$ and $\Delta_2'$ (the masses and hence the $\tht'$ follow from this
because $m=|\Delta|$), and is
\eqn\Qiifinal{\eqalign{e^{i\pi f_i}
{i\beta\over 2(2\pi)^2}\int\int d\tht_1 &d\tht_2
\left(({\partial\over\partial\tht_1}+
{\partial\over\partial\tht_2})(k_1+k_2)\right)\cr
&{\partial\over\partial\tht_1} Im \ln {\sinh({{\tht_2-\tht_1}\over
2}+\mu)\over \sinh({{\tht_2'-\tht_1'}\over 2}+\mu')}
e^{-\beta (m_1\cosh\tht_1+m_2\cosh\tht_2)}.\cr}}
The result simplifies in the case of elastic scattering, where the masses of
the particles do not change.  (Forward elastic scattering is the only allowed
process in integrable theories.) For forward elastic scattering, the kinematic
prefactor in \Qiifinal\ is $2m_1\cosh\tht_1+2m_2\cosh\tht_2$. Moreover, for
forward or backward elastic scattering, one has
$$\tht_1-\tht_2=\tht_1'-\tht_2',\qquad \mu'=-\mu,$$
showing clearly that the elastic two-particle contribution to the index
vanishes only when $\mu= 0, i\pi/2.$

For the two-particle contribution from all the reduced multiplets, we sum
\Qiifinal\ over all pairs.  We have thus
seen that two-kink contribution to the index
from kinks belonging to reduced supersymmetry
multiplets is non-vanishing and easily computable.

The results in this section can be compared with the infra-red limit of the
index as obtained from the $tt^*$ differential equations discussed in the
previous section.  It is non-trivial that they agree, but they must.
This is being investigated numerically in some examples of $N$=2 theories
which are not integrable \bill . In the next sections we will focus on
integrable theories since then we can obtain, via the exact $S$ matrix and
thermodynamics, exact integral equations for the index which can be compared
with the $tt^*$ equations along the entire renormalization group flow.

\newsec{\F\ in integrable theories}
In this section we will show how to compute the index for an integrable
theory, using the exact $S$-matrix. Integrable theories in two dimensions have
been under intensive investigation recently.  These theories are characterized
by the existence of infinitely many conserved charges, which essentially
allows one to solve these theories explicitly.  In particular the scatterings
are purely elastic; the particles behave as if they are free particles and as
they pass through each other they just pick up phases (modulo change of
internal indices).  The multi-particle $S$-matrix factorizes into two-body
$S$-matrices and these are often completely determined by the symmetries of
the theory (plus some minimality assumptions which can be verified \tbaref ).
The factorizability and elasticity of the $S$-matrix in an integrable theory
implies that we can assign rapidities (momenta) to individual particles even
in multi-particle configurations.  In particular the total energy of
multi-particle configurations is simply the sum of the individual ones. The
only effect of the interaction is to shift the
density of allowed states.  This is
an ideal situation for computation of our index; the non-trivial part of our
index \ivdiml\ results precisely from a discrepancy between densities of
states within a non-reduced supersymmetry multiplet.

It is clear that we can in principle continue the analysis of the previous
section, using the exact $S$-matrix of an integrable theory, to calculate
higher-order corrections in the infra-red expansion.  In fact, we can do {\it
much} better. There is a trick (known as the {\it Thermodynamic Bethe Ansatz}
\tbaref ) which allows us to compute the index exactly along the entire
renormalization group flow, even in the {\it ultra-violet} limit!  The idea is
to not fix the number of solitons, but to consider a thermodynamic ensemble of
them.  We then minimize the free energy in the ensemble.  As we will review,
this allows us to calculate exactly, i.e. non-perturbatively, \F\ in an
integrable theory with a known exact $S$-matrix.  In fact it is no more
difficult to compute the more general quantity $\log \Tr(e^{i\alpha
F}e^{-\beta H})$.  This allows us to test our claim that while this quantity
depends on the $D$-terms its first derivative with respect to $\alpha$ at
$\alpha =\pi$ is independent of the $D$-terms.  Even more generally, let us
consider the free energy ${\cal F}_{\mu _a} (\beta )$ with chemical potentials
$\mu _a$ for the various conserved species labels
\eqn\Emu{-\beta {\cal F}_{\mu _a}(\beta )=\ln \Tr (e^{\beta \sum _a \mu
_aN_a}e^{-\beta H}),}
where $N_a$ is the number operator for conserved species $a$.  Using the exact
$S$-matrix we can obtain an exact expression for $\beta {\cal F}_{\mu _a}(\beta
)$ by finding the minimum value of $\beta E-S-\beta \sum_a \mu _a N_a$
in the space of all states, where $S$ is the entropy.  Choosing the chemical
potentials in \Emu\ such that $\beta\sum _a\mu _aN_a=i\alpha F$, we thereby
obtain $\Tr(e^{i\alpha F}e^{-\beta H})$.

Since the $S$-matrix of an integrable theory preserves rapidities and some set
of species labels $a$, a general, multi-soliton state can be characterized by
a collection of
distributions $\rho _a(\t )$ of rapidities occupied by the various solitons in
the multi-soliton state.  In particular, the energy of this state is given by
\eqn\Eis{E=\sum _a \int d\t \rho _a (\t )m_a\cosh \t .}
To do thermodynamics we need to calculate
the entropy $S$ and so we need to know
 the
distributions $P_a(\t )$ of available levels as well as the above
distributions $\rho _a(\t )$ of occupied levels.  In particular
\eqn\Sis{S=\sum _a\int d\t \ P_a\log P_a-\rho _a\log \rho _a -(P_a-\rho _a)\log
(P_a-\rho _a),}
corresponding to one particle allowed per level.

Using the exact, factorizable $S$-matrix it can be found that the distributions
of available levels are given in terms of the distributions of occupied levels
in the general manner:
\eqn\Pdist{2\pi P_a(\t )=m_aL\cosh \t +\sum _b\int d \t '\rho _b(\t ')
\phi _{ab}(\t-\t ').}
 The $m_aL\cosh \t $ term in \Pdist\ is the usual
density of available states
for a free particle, the $\phi _{ab}$ reflect the interaction with the
other particles, as given by the exact $S$-matrix.
If the $S$-matrix is diagonal,
with species $a$ and $b$ scattering with the phase shift $S_{ab}$, the
interaction is seen to be given by $\phi _{ab}(\t )=-i\partial \log S_{ab}(\t
)/\partial \t$.  For non-diagonal S-matrices such as our $N$=2 S-matrices, it
is generally difficult to obtain the $\phi
_{ab}$ from the $S$-matrix; one needs
to find the eigenvalues of the multi-particle transfer matrices.

Now we minimize $\beta E-S-i\beta \sum _a\mu _a N_a$,
expressed in terms of the above distributions, with respect to the $\rho _a(\t
)$ subject to the constraints \Pdist .  Defining the  quantities $\e _a(\t )$
by
\eqn\pe{{\r _a(\t )\over P _a(\t )}={e^{\beta \mu _a-\e _a(\t )}\over 1+
e^{\beta \mu _a-\e _a(\t)}},}
it is seen that the free energy is given by
\eqn\fe{\log \Tr (e^{\beta\sum _a \mu _a N_a}e^{-\beta H})=-\sum _a
m_aL\int {d\t \over 2\pi }
\cosh \t \ln (1+e^{\beta \mu _a-\e _a(\t)}),}
where the $\e _a (\t )$ are obtained as the solutions to the coupled integral
equations:
\eqn\TBA{\e _a(\t )=m_a\beta\cosh (\t )-\sum _b \int {d\t ^{\p}\over 2\pi}\phi
_{ab}(\t -\t ^{\p})\ln(1+ e^{\beta\mu _b-\e _b(\t ^{\p})}).}
These are the thermodynamic Bethe ansatz \tbaref\ integral equations with
chemical potentials \rKMi.\foot{We note that it is straightforward to
rederive the TBA equations with our fixed boundary conditions instead of the
usual periodic one. The result is the same.} Our interest is in the case where
the chemical potentials are chosen such that $\beta \sum _a\mu _a N_a=i\alpha
F$.  The expression \fe\ was obtained by summing over all boundary conditions
at spatial infinity.  The different eigenvalues of our matrix trace can be
obtained from this expression by inserting appropriate additional chemical
potentials.  Examples will be discussed in the following section.

\newsec{Examples}

A generic $N$=2 theory will not, of course, be integrable.  Nevertheless, our
index for such a theory can be obtained by solving the differential equations
discussed in the $tt^*$ section of this paper.  We would like, however, to
compare the computation of the index from the $tt^*$ differential equations
with the computation from the thermodynamic integral equations.  We will thus
restrict our examples to integrable theories for which the exact $S$-matrix is
known (or conjectured). Examples of such theories are discussed in
\refs{\pk,\pkii}.

We will focus on integrable theories with spontaneously broken \Zn\ symmetry.
For every $n$ there are a wide variety of such examples, including
perturbations of $N$=2 minimal models and Kazama-Suzuki models, supersymmetric
$CP^{n-1}$ sigma models, and $N$=2 affine toda
theories.  For a given value of $n$, the $tt^*$ differential equations and the
TBA integral equations for all these \Zn\ integrable theories are found to be
essentially the same, the only variation being in the boundary conditions.  We
will first consider several examples of ${\bf Z}_2$ theories, namely ordinary
$N$=0 sine-Gordon at the particular coupling where it is $N$=2 supersymmetric,
$N$=2 sine-Gordon, $N$=2 minimal models perturbed by the least relevant
perturbation, and the supersymmetric $CP^1$ sigma model.  The indices for all
of these theories are obtained from the same differential equation, Painleve
III.  They span {\it all} the possible regular boundary conditions.  The TBA
integral equations also exhibit this fact in a non-trivial guise.  We next
discuss the more general \Zn-type integrable theories starting with the
simplest such theory, the $A_n$ $N$=2 minimal model perturbed by the most
relevant operator.  We finally discuss how to modify the equations in order to
determine the index for the other \Zn-type theories.

For the most part, we will consider $N$=2 theories which can be described by a
Landau-Ginzburg action\foot{The existence
of Landau-Ginzburg description seems to apply
also to non-supersymmetric and $N=1$ supersymmetric theories \rLG .}.
Such an action is of the form \rVW
$$\int K(X_i,\bar X_i )+ \int W(X_i )+\int {\bar W(\bar X_i)}$$
where the superfields $X_i$ are chiral (in supersymmetric sense, i.e.
annihilated by $D^+, \bar D^+$) fields, $W$ is the superpotential and is
integrated over half the superspace, and $K$ gives the kinetic terms (the
$D$-term) and is integrated over the full superspace.  Using topological
techniques one can prove that the chiral ring of this theory is exactly
characterized by $W$ \tlg .  In particular the chiral fields of the theory are
all products of the fields $X_i$ modulo setting to zero $\partial _i W$.  The
chiral ring structure constants entering in \curv\ are obtained by simply
multiplying the various products of $X_i$ together and imposing the relations
$\partial _i W$=0.

The physical potential for the theory is
$$V=K^{i\bar\jmath}\partial_i W\partial_{\bar\jmath}\bar W$$
where $K^{i\bar\jmath}$ is the inverse of the Kahler metric
$\partial_i\partial_{\bar\jmath}K$.  The vacua $a$ of the theory are thus in
one-to-one correspondence with the critical points of $W$.  The kinks $k_{a
b}$ are the finite energy solutions to the equations of motion connecting the
$a$ and $b$ vacua: $X(\sigma =-\infty )=X^{(a )}$, $X(\sigma = +\infty)=X^{(b
)}$ (as discussed in \LGSol , not all such kinks are to be regarded as
fundamental solitons).  The central term in the $N$=2 algebra \sual\ is given
simply in terms of the superpotential by $\Delta =2\Delta W\equiv 2[W(X(\sigma
=+\infty))-W(X(\sigma =-\infty))]$.  The mass of the $(u,d)$ soliton doublet
representation is thus given simply in terms of the superpotential by
$m=2|\Delta W|$.

The fractional fermion number in the soliton sector is also given simply in
terms of the superpotential by a spectral flow argument \pkii\ or by adiabatic
or index theorem techniques \refs{\rGW ,\rNS}.  The result is that the $(u,d)$
soliton doublet has the fermion numbers $(f,f-1)$ where
\eqn\fract{f=-{1\over 2\pi}(Im \ln \det(\partial _i\partial _jW(X)))\big
|^{\sigma=+\infty}_{\sigma=-\infty}.}
In all of the integrable theories we consider, the entire spectrum consists of
such soliton doublets saturating the  bound.  There are other
integrable $N$=2 theories, for example the theories with superpotential
$W=x^{n+1}/(n+1)-\lambda x^2$ \refs{\LGSol,\toda} , for which this is not the
case; exact $S$-matrices for these and many
other integrable $N$=2 theories have
been recently discussed in \lnw .

\subsec{$N=0$ sine-Gordon at the $N$=2 point}

Ordinary $N=0$ sine-Gordon theory is $N$=2 supersymmetric at a particular
coupling \ref\sgs{G. Waterson, Phys. Lett. B171 (1986) 77}.  In a manifestly
$N$=2 supersymmetric setup, this point is described by a Landau-Ginzburg
superpotential
\eqn\xcubed{W=\lambda ({X^3\over 3}-X),}
with some suitable choice of $K$ \refs{\rVW}. The vacua $|a\>$ are at $X=\pm
1$.  Our matrix index $Q$ has eigenvalues $Q(z)$ and $-Q(z)$, where $z=m\beta
$ and $m=2|W(X=1)-W(X=-1)|={8\lambda \over 3} $.  We will first use the $tt^*$
equations to find an exact expression for our index in terms of a famous
differential equation.

The $tt^*$ analysis of this theory was discussed at length in \cv .  The
result is that the metric on the space of ground states, in the basis spanned
by $1,X$, is given by $g=e^{\sigma _3u(z)/2}$. Using \curv\ with the chiral
ring $X^2=1$, it follows that $u(z)$ satisfies the radial sinh-Gordon equation
\eqn\piii{   {d^2u \over dz^2}+{1\over
z}{du\over dz}=\sinh u.}
The radial sinh-Gordon equation is a special case of Painleve III. {}From
relation \final\ it follows that the index $Q(z)$ is given by
\eqn\piiiQ{Q(z)={1\over 2}z{d\over dz}u(z).}
If we wish, we could eliminate $u(z)$ from these equations and write a
differential equation for $Q(z)$
\eqn\Qpiii{Q ''-z^{-1}Q'=Q\sqrt{4z^{-2}Q'^2+1}.}

The solutions $u(z)$ to \piii\ behave for $z\rightarrow 0$, i.e. the
ultra-violet or conformal limit, as
\eqn\piiibc{\eqalign{u(z;r)&\sim r\log {z\over 2} + s +O(z^{2-|r|})
\qquad \hbox{with}\ |r|<2\cr &\sim \pm {\rm log}\ {z\over 2}\pm {\rm log}[-
({\rm log}({z\over 4}+C)]+O(z^4{\rm log}^2z) \qquad \hbox{for}\ |r|=2}}
where
$$e^{s/2}={\Gamma ({1\over 2}-{r\over 4})\over
2^r\Gamma ({1\over 2}+{r\over 4}) }$$
and where $C$ is the Euler constant, and $r$ is just a parameter to label the
boundary condition.  For our theory regularity requires $r$=2/3 \cv .  {}From
$\piiibc$ it follows that $\pm Q(z=0)=\pm 1/3$.  This is to be expected, since
in this limit the eigenvalues of the $Q$ matrix index are the left plus the
right $U(1)$ charge of the Ramond ground states.

We will now obtain integral equations for the function $Q(z)$ \pk , using the
exact $S$-matrix obtained in \refs{\rZandZ, \rBL} .  The one particle spectrum
of this theory consists of a soliton reduced multiplet $(u,d)$, with mass
$m=2|\Delta W|$, and fermion numbers obtained from \fract\ to be $(1/2,-1/2)$.
The soliton connects either vacuum with the other one.  A multi-soliton state
can be characterized by a distribution $\rho _1(\t )$ of rapidities occupied
by the solitons.  Because the $u$ and $d$ solitons do not scatter diagonally,
we can not assign individual distributions for $u$ type and $d$ type solitons.
Instead, there are two additional distributions $\rho _l(\t )$, $l=0,\bar 0$,
which encode the way in which the solitons are distributed as $u$ or $d$
solitons.  The distributions $\rho _l(\t )$ arose in \pk\ in obtaining the
eigenvectors and eigenvalues of the multi-particle transfer matrices.  We
correspondingly have distributions $P_1(\t )$ and $P_l(\t )$ for the density
of states.  As discussed in \pk , these densities satisfy relations of the
type \Pdist , described by the diagram
\bigskip
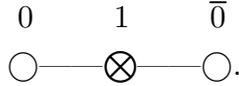
\centerline{\hbox{\rlap{\raise15pt\hbox{$\ \,\, 0\hskip.89cm 1\hskip.89cm
\bar 0$}}}
$\bigcirc$------{\hbox{$\bigotimes$}}------$\bigcirc .$}
\bigskip
\noindent
The nodes in this diagram correspond, as labeled, to the species in \Pdist .
The nodes for species 0 and $\bar 0$ are open to signify that these species
have $m_a$=0 in the equations \Pdist (arising from the fact that these species
are not physical particles but, rather, account for the additional $u$ and $d$
degree of freedom); the $\otimes$ node has mass $m_1=2|\Delta W|$.  The
functions $\phi _{ab}$ in \Pdist\ are given by $\phi _{ab}(\t )=(\cosh \t
)^{-1}l_{ab}$, where $l_{ab}$ is the incidence matrix for the figure, i.e. it
is one when species $a$ and $b$ are connected by a line and zero otherwise
(and $l_{aa}$=0).

The remaining ingredients required in equation \TBA\ are the chemical
potentials. These are chosen so that $\beta\sum_a\mu_aN_a=i\alpha F$.  Thus we
need to express $F$ in terms of the above densities.  For a state with a total
number $k$ of $u$ solitons and a total number $N-k$ of $d$ solitons, these
distributions are defined to satisfy
$$\int d\t \rho _1(\t )=N \quad \hbox{and}\quad
\int d\t (P_{0}-\rho _0+\rho _{\bar 0})=k.$$
The fermion number of such a state is $k-(N/2)$ and so using the above
equations along with \Pdist\ we find
\eqn\fis{F=\int d\t (\rho _{\bar 0}(\t )-\rho _0 (\t )).}

We are ready to use \fe\ and \TBA\ to obtain exact integral equations for
\break$\ln\Tr e^{i\alpha F}e^{-\beta H}$.
The $m_a$ and functions $\phi _{ab}(\t )$
in these equations are as given above and, using \fis , the chemical
potentials should be taken to be $\beta \mu _1=0$ and $\beta \mu _{0}=-\beta
\mu _{\bar 0}=i\alpha$.  First note that at $\alpha=\pi$ (Witten's index) the
equations \TBA\ are solved by
$$e^{-\e _1(\t)}=\e _{0}(\t )=\e _{\bar 0}(\t )=0$$
for all $\tht$. {}From \fe\ is is then seen that $L^{-1}\log \Tr
(-1)^Fe^{-\beta H}=0$ when $L\rightarrow
\infty$, as expected.

To move slightly away from Witten's index, take $\alpha =\pi +h$ with $h$
small and keep terms only up to $O(h)$.  The equations \TBA\ are then solved
by $e^{-\e_1 (\t )}=he^{-A(\t)}$, and $\e _0(\t )=\e _{\bar 0}(\t )=hB(\t )$,
where
\eqn\piiiint{\eqalign{
A(\t ;z)&=z\cosh \t -\int {d\t '\over 2\pi }{1\over \cosh (\t -
\t ')}\ln (1+B^2(\t';z))\cr
B(\t ;z)&=-\int{d\t '\over 2\pi }{1\over \cosh (\t -\t ')}e^{-A(\t ';z)}.}}
Using \fe\ and \indd , the $Q$-matrix eigenvalues are given by $\pm
Q(z=m\beta)$ where, in terms of the solution to the above equations
\eqn\piiiF{Q(z)= z\int {d\t \over 2\pi}\cosh \t e^{-A(\t ;z)}.}

We have two exact expressions for the index $Q(z)$, the differential equation
\piii\ and \piiiQ\ (or \Qpiii) obtained from $tt^*$ considerations, and
 the integral equations \piiiint\ and \piiiF\ obtained from $S$-matrix and
thermodynamic considerations.  These expressions must agree!  We know of no
way, however, to show directly from the equations that this is the case.
Physics has proven a highly non-trivial statement about the above equations.
A check is that the ultra-violet limit for the index using the two different
equations give exactly the same result $Q(z=0)=1/3$.  Also, using results from
\wu\ concerning the Painleve III differential equation, it can be seen (after
some algebra) that the function $Q(z)$ has an expansion
\eqn\mcwform{Q(z)=-z{d\over dz}\sum _{n=0}^{\infty}
{2(2\cos({\pi \over 2}(2-r)) )^{2n+1}\over 2n+1}\int \prod
_{i=1}^{2n+1}{e^{-z\cosh \theta}\over \cosh ({\theta _i-\theta _{i+1}\over
2})}{d\theta _i\over 4\pi}}
($\theta _{2n+2}\equiv \theta _1$) and, for later use, we have restored the
boundary condition parameter $r$ from \piiibc ; for the present example
$r=2/3$. \foot{A similar expression arose in the computation of Ising-model
form factors \ref\card{J. Cardy and G.  Mussardo, Nucl. Phys. B 340 (1990)
387}.} This is the type of infra-red (large $z$) expansion which we would
expect for our index; the first term is the usual Bessel function.  Also, only
odd numbers of solitons contribute because only then are the vacua at spatial
infinity different. It is easy to see that the two-particle contribution
computed in section 4 vanishes, because $\mu=i\pi/2$ here.  By expanding out
the integral equations \piiiF\ and \piiiint\ it is easily seen that the
one-soliton and the three-soliton contribution agree with the $n$=0 and $n$=1
terms in the PIII expression \mcwform ; after that the comparison becomes more
difficult to check directly.  We have numerically verified (to real precision)
that the function $Q(z)$ obtained from \piii\ and \piiiQ\ does, indeed, agree
with that obtained from \piiiF\ and \piiiint. It would be interesting to see
how difficult it is to find a direct mathematical argument to verify this.  In
particular, would we have to re-invent the physics argument, in disguise, to
prove their equality?!

\subsec{$N$=2 Super sine-Gordon}

As our next example, we consider the $N$=2 super sine-Gordon theory
given by the Lagrangian
\eqn\laga{\int d^2z d^4 \t X\bar X +{m\over 4}(\int d^2z d^2\t \cos g X +h.c.)}
The coupling $g$, by a redefinition of $X$ and $\bar X$, can be taken to be
real.  Because our index is independent of the $D$-term, it must, in fact, be
independent of the coupling $g$ since, by rescaling the chiral fields, $g$ can
be eliminated from the $F$-term and put into the $D$-term.  The general
quantity $Z(\alpha,\beta)={\rm Tr} e^{i\alpha F}e^{-\beta H}$ can be
calculated via the TBA equations since this is an integrable theory.
$Z(\alpha,\beta)$ depends on $g$ as a sign of its dependence on the $D$-term.
We will show that our index, the first derivative with respect to $\alpha$ at
$\alpha =\pi$, is independent of $g$ as expected by our general arguments in
sects. 2 and 3.

The vacua of the $N$=2 sine-Gordon theory are the points $gX_a=a\pi$ for $a\in
{\bf Z}$.  We thus have an infinite number of possible vacua for the boundary
conditions at $\sigma =\pm \infty$.  Because of the symmetry $gX\rightarrow
gX+\pi$, a configuration with vacuum $X_{a}$ to the left and vacuum $X_b$ to
the right is equivalent to one with vacuum $X_{a+n}$ to the left and vacuum
$X_{b+n}$ to the right.  Consider the contribution $Q_{ab}$ to our index from
a fixed boundary condition $(ab)$.  Then, this symmetry implies that
$Q_{ab}=Q_{a+n,b+n}$.  The eigenvalues of a matrix $M_{ij}$ whose entries
depend only on $i-j$ are easily found by Fourier transform.  The eigenvalues
$Q(\Theta )$ are parametrized by an angle $\Theta$ and given by
\eqn\qzT{Q(z; \Theta )=\sum _{l=-\infty}^{\infty} e^{i\Theta l}Q_{i,i+l}(z).}
In other words, we weight configurations by $e^{i\Theta T}$ where $T$ is the
topological charge.

Using the results of \cv\ and \final, the index eigenvalues $Q(z;\Theta )$ are
all solutions of the same PIII differential equation \piii\ obtained in the
last example, where only the boundary condition \piiibc\ depends on $\Theta$.
Regularity requires \cv\ the solution $Q(z, \Theta )$ to behave (with
$z=m\beta$) as \piiibc\ with
\eqn\rT{r(\Theta )=2(1-{2\Theta \over \pi})\qquad
\hbox{for}\ \ 0\leq \Theta \leq \pi,}
with $r(\Theta )$ defined outside this interval by $r(\Theta +\pi )=-r(\Theta
)$, a consequence of the fact that, in \qzT , only configurations with odd $l$
contribute. By varying $\Theta$, we obtain all regular solutions of Painleve
III.

Now we come to the analysis of this theory from the viewpoint of the TBA.  The
close connection between the index in this example and that in the previous
example can also be seen from the exact $S$-matrix and associated integral
equations.  The important point is that every sine-Gordon soliton is the same
$(u,d)$ supermultiplet, with fermion numbers $(1/2,-1/2)$, seen in the
previous example.  The (conjectured \Kob ) $S$-matrix for the $N$=2 sine
Gordon theory is simply the tensor product of the $S$-matrix for the theory
considered in the previous example with the $S$-matrix for the $N$=0
sine-Gordon \rZandZ\ at coupling $g^{bare}_{N=0}=g$.  The TBA integral
equations for this theory are obtained by combining those of the previous
section with the TBA system for $N$=0 sine-Gordon\pkii.  The TBA system of
integral equations for $N$=0 sine-Gordon at generic coupling $g$ is of the
usual form \TBA\ but with an infinite number of species $a$ and a complicated
set of $\phi _{ab}(\t )$.  For the sake of brevity we will thus focus on a
nice set of couplings, $g^2=8\pi s$ for $s$ a positive integer $s\geq 2$,
where the equations simplify\fs.  Of course, our index is independent of the
coupling $g$ so we can work with any coupling we please.  We will verify this
fact, though the more general quantity \parti\ does depend on $s$.

At the coupling $g^2=8\pi s$, as discussed in \pkii , we obtain a TBA system
of coupled integral equations of the usual form \TBA\ for $s+3$ functions $\e
_a(\t )$.  The masses $m_a$ and $\phi _{ab}(\t )$ entering in the equations
\TBA\ for this theory are described by the figure
\bigskip
\noindent
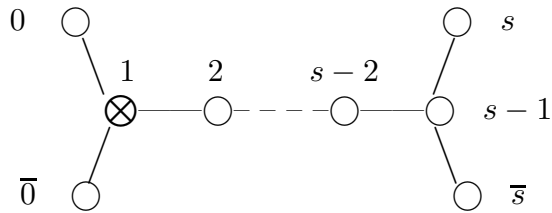
\centerline{
\hbox{\rlap{\raise28pt\hbox{$0\hskip.3cm\bigcirc\hskip
3.8cm\bigcirc\hskip.3cm s$}}
\rlap{\lower27pt\hbox{$\bar 0\hskip.3cm\bigcirc\hskip
3.8cm\bigcirc\hskip.3cm \bar{s}$}}
\rlap{\raise14pt\hbox{$\hskip.55cm\Big\backslash\hskip3.6cm\Big/$}}
\rlap{\lower14pt\hbox{$\hskip.5cm\Big/\hskip3.55cm\Big\backslash$}}
\rlap{\raise12pt\hbox{$\hskip.75cm 1\hskip .80cm 2\hskip .95cm s-2$}}
$\hskip.45cm${\hbox{$\bigotimes$}}------$\bigcirc$-- --
-- --$\bigcirc$------$\bigcirc\hskip.3cm s-1$ }}
\bigskip

\noindent
where every node in the diagram corresponds to a species in the equations
\TBA. The $\otimes$ node, labelled by 1, corresponds to the soliton; its mass
in the equations \TBA\ is that of the soliton.  The other species have open
nodes to signify that they have $m_a$=0 in the equations \TBA. Again, the role
of these additional species is to account for the additional degrees of
freedom (i.e.\ $u$ or $d$, and which vacua they connect).  As in the previous
example, the $\phi _{ab}( \t)=(\cosh \t )^{-1}l_{ab}$ where $l_{ab}$ is the
incidence matrix for the above figure.  The massive node and the nodes
labelled $0$ and $\bar 0$ come from the $N$=2 part of the $S$-matrix; they
correspond precisely to the species in the previous example.  The massive node
connected to the $s$ open nodes are the TBA species for $N$=0 sine-Gordon at
coupling $(g^{bare}_{N=0})^2=8\pi s$.  This result is obtained by using a
technique known as the algebraic Bethe ansatz to find the eigenvalues of the
multi-soliton sine-Gordon transfer matrices (see the appendix of \fs ).  The
TBA system for the tensor-product $S$-matrix is obtained by joining the two
component TBA systems at the massive node as described by the above figure.

As in the last example, the fermion number is given by
$$ F=\int d\t (\rho _{\bar 0}(\t )-\rho _0 (\t )),$$
whereas the sine-Gordon topological charge (number of solitons minus anti
-solitons) is given by \fs\
\eqn\tis{T=s\int d\t (\rho _{\bar{s}}(\t )-\rho _{s}(\t)),}
where the species labels are as given in the above figure.  By introducing
chemical potentials $\beta \mu _{\bar 0}=-\beta \mu _0 =i\alpha$ and $\beta
\mu _{\bar{s}}=-\beta \mu _{s}=is\Theta$, with the other chemical potentials
zero, the equations \TBA\ and \fe\ provide integral equations to compute $\log
\Tr e^{i\alpha F}e^{i\Theta T}e^{-\beta H}$ exactly.  For generic $\alpha$,
the integral equations, in particular the number of functions $\e _a (\t )$,
clearly depend on $s$.  In the infra-red expansion, one sees that the
solutions of these equations are in fact different.  Thus $\ln\Tr(e^{i\alpha
F}e^{i\Theta T}e^{-\beta H})$ depends on the coupling $g^2=8\pi s$, as
expected.

At $\alpha$=$\pi$, the solution of the equations \TBA\ described by the above
figure with the above chemical potentials is given by $\e _a (\t )$
independent of $\t$: $\e _0 (\t )$=$\e _{\bar 0}(\t )$=$e^{-\e _1(\t )}$=0,
and
\eqn\eais{\eqalign{&e^{-\e _a(\t )}=({\sin  a\Theta \over \sin  \Theta})^{2}
-1 \qquad   \hbox{for}\ a=2, \dots ,s-1\cr
&e^{-\e _{s}}=e^{-\e _{\bar{s}}}={\sin (s-1)\Theta \over \sin \Theta }.}}
As expected by supersymmetry, \fe\ gives $L^{-1}\log \Tr (-1)^Fe^{-\beta H}
=0$ in the $\L \rightarrow \infty$ limit.

We now move slightly away from Witten's index.  At $\alpha =\pi +h$ with $h$
small, the solution of the equations \TBA\ is given by $\e _0 (\t )$=$\e
_{\bar 0}(\t )$=$h B(\t )$, and $e^{-\e _1(\t )}$=$h A (\t )$, where the
functions $A(\t )$ and $ B (\t )$ obey the equations
\eqn\pintt{\eqalign{
A(\t)&=z \cosh \t -\ln (2\cos \Theta )-\int {d\t '\over 2\pi }{1\over \cosh
(\t -\t ')}\ln (1+B^2(\t'))\cr B(\t )&=-\int{d\t '\over 2\pi }{1\over \cosh
(\t -\t ')}e^{-A(\t ')}.}}
The other $\e _{a}(\t )$ are (to lowest order in $h$) still given by the
constants \eais .  Using \fe\ and \indd\ we thus see that the $Q$ matrix
eigenvalues are given by $ Q(z=m\beta ;\Theta )$ where
\eqn\piiiFt{Q(z; \Theta )= z\int {d\t \over 2\pi}\cosh \t e^{-A(\t ;z;\Theta )
},}
with $A(\t ;z;\Theta )$ obtained by solving \pintt .

As expected, our index is independent of the coupling $g$ (i.e. $s$).
One can check by for example studying the IR expansion of the solutions
to the full TBA equations that all the other quantities $I_l$ for $l>1$
{\it do} depend on $g$ and are thus dependent on $D$-terms, in accord
with the arguments of section 2.
 In fact, the entire $N$=0 part of the theory has dropped out of the integral
equations, leaving just the constant $\ln 2\cos\Theta $ in \pintt, which
resulted from the constants \eais .  This constant piece is the only
reflection of the $N$=0 sine-Gordon structure.  The close connections between
these integral equations and those computed in the previous example was also
to be expected from the $tt^*$ considerations; the $\ln (2\cos \Theta)$ term
in \pintt\ specifies the boundary conditions in the Painleve III equation.
Again, though \pintt\ and \piiiFt\ have no obvious connection to the radial
sinh-Gordon (or Painleve III) differential equation \piii\ and \piiiQ ,
physics proves that
the regular solutions are the same.

It is easily seen that \pintt\ and \piiiFt\ lead to an expansion of the form
\eqn\nsoliton{Q(z; \Theta )=\sum _{n=0}^{\infty}(2\cos
\Theta )^{2n+1}A_{2n+1}(z).}
$A_{2n+1}$ is the contribution from the $(2n+1)$-soliton sector, and is
independent of $\Theta$.\foot{This leads to an amusing intuition for our
index.  Write this result as
$$Q(z;\Theta)=\sum _{n=0}^\infty A_{2n+1}(z)
(e^{i\Theta}+e^{-i\Theta})^{2n+1},$$
and compare with \qzT .  The factor $e^{i\Theta}$ counts the solitons
connecting vacuum $i$ to vacuum $i+1$ (topological charge 1) and the factor
$e^{-i\Theta}$ counts anti-solitons connecting vacuum $i$ to vacuum $i-1$
(topological charge $-1$). The fact that they are all weighted by the same
factor means that we can obtain the index by
weighting each $2n+1$-soliton configuration by $A_{2n+1}$.
This intuition only applies to the computation of our index, and
not to the other thermodynamic quantities.} In the
large $m\beta $ limit, $A_{2n+1}$ is $O(e^{-(2n+1)m\beta})$). We can compare
this result with \mcwform .  We immediately arrive at the expression \rT\
relating the boundary condition $r$ on the PIII differential equation to our
parameter $\Theta$.  This is a further confirmation that the TBA solution is
related to PIII solution, as in both cases the dependence on PIII boundary
conditions $r$ or equivalently $\Theta$ has the same structure.  As a further
check, the UV limit $z\rightarrow 0$ of \piiiFt\ and \pintt\ is obtained
(using results from \rPF\ for taking the UV limit of TBA systems with
imaginary chemical potentials) to be
\eqn\piiiuv{Q(z=0; \Theta )=(1-{2\Theta \over \pi}),}
for $0< \Theta < \pi$, in agreement with \piii\ and \piiiQ\ with the boundary
condition specified by \piiibc\ and \rT .

At $\Theta$=0 mod $\pi$ we have to be more careful in evaluating the UV limit
of the equations \piiiFt\ and \pintt .  Exactly as in \rAlZiv , there is a log
piece in the ultra-violet limit.  This log piece agrees with the expression
\piiibc\ for the Painleve III solution at $r=\pm 2$.

\subsec{$N$=2 minimal models with least-relevant perturbation}

An $N$=2 minimal model remains integrable when perturbed by its least-relevant
operator \refs{\toda,\rMW}.  The effective Landau-Ginzburg superpotential for
the perturbed theory is identified to be
a Chebyshev polynomial (as conjectured in \cv\ and
confirmed in \pk ); e.g. for the perturbed
$A_{k+1}$ theory the superpotential is $W_{k}(X=2\cos \t )=(2/k+2)\cos
(k+2)\t$, expressed as a polynomial in $X$.  This is the least relevant
perturbation of the conformal field theory $W=X^{k+2}/k+2$ in the flat
direction of \dvv .  The chiral ring of this theory yields the $SU(2)_k$
fusion rules \rGep .

The $tt^*$ computation of our index here is a simple application of the
results from the previous example.  The reason is that if we change variables
$(k+2)\t \rightarrow Y$, the above Chebyshev polynomial becomes the
superpotential for $N$=2 sine-Gordon.  The eigenvalues of the index are thus
obtained to be
\eqn\chebind{Q(z=m\beta; \Theta ={\pi n\over k+2}),}
for $n=1, \dots, k+1$, where $Q(z;\Theta )$ is the function discussed in the
previous example.

We now discuss the TBA calculation of the index using the exact $S$-matrix for
these theories.  The vacua of the $W_k (X)$ Chebyshev theory are at $X^{(n)}
=2\cos {(\pi n/ k+2)}$, for $n=1,\dots ,k+1$.  The solitons are $N$=2
Bogomolnyi doublets $(u_j, d_j)$ connecting vacuum $X^{(j)}$ with vacuum
$X^{(j+1)}$, for $j=1, \dots k$, each identical to that of the example in sect
6.1; they all have the same mass and fermion numbers $(\half, -\half )$.  The
structure of $k+1$ vacua on a line is that of the $k$-th RSOS theory \RSOS,
which describes the $N$=0 minimal models with least-relevant perturbation.
The $N$=2 Chebyshev $S$-matrix is a direct product of this $N$=0 RSOS
$S$-matrix with the $N$=2 $S$-matrix discussed in sect. 6.1, just as the $N$=2
sine-Gordon $S$-matrix of the previous subsection was the tensor product of
the $N$=0 sine-Gordon $S$-matrix with the $N$=2 $S$-matrix of sect. 6.1.

There is a well-known reduction from $N$=0 sine-Gordon at a particular
coupling to the $k$-th RSOS theory.  This same reduction can be used to obtain
our $N$=2 Chebyshev theories from the $N$=2 sine-Gordon theory; the common
$N$=2 structure just goes along for the ride.  We start with the $N$=2
sine-Gordon TBA equations appropriate for $N$=2 sine-Gordon coupling
$g^2=8\pi (k+2)$.  The equations are described by the diagram of the previous
subsection, with $s=k+2$.  The reduction of this to our Chebyshev theory
requires taking $e^{\beta \mu _{s}}=e^{\beta \mu _{\bar {s}}}=-1$ \fs.  This
reduction simply eliminates the nodes labelled $s$, $\bar s$, and $s-1$ from
the diagram in the previous subsection, leaving the TBA system discussed in
\pk .

The solution for this TBA system is the sine-Gordon solution $Q(z;\Theta =\pi
/(k+2))$, giving the largest eigenvalue of the $Q$-matrix for the Chebyshev
theory.  The remaining eigenvalues of the matrix index are given by $ Q(z,
\Theta =\pi n/(k+2))$, for $n=1, \dots, k+1$, and can be seen as other
branches of this solution.\foot{The resulting factors of $2\cos n
\pi/(k+2)\equiv\l_n$ in \nsoliton\ are the eigenvalues of the RSOS soliton
incidence matrix. The number of $N$-soliton configurations can be expressed
(for any boundary condition) as $\sum _nc_n\l_n^N$ with $N$ independent
$c_n$.  This is in accordance
with the intuition of the footnote following \nsoliton.} It follows from
\piiiFt\ that the non-zero eigenvalues come in pairs of opposite sign, as they
should.  As another check, note that in the ultra-violet limit we obtain
\piiiuv\ $Q(z=0, \Theta =\pi n/(k+2))$= $1-(2n/k+2)$; these are the correct
expressions for the left plus right chiral $U(1)$ charges for the Ramond
ground states of the conformal theory obtained in the ultra-violet limit.

\subsec{Supersymmetric $CP^1$ sigma model}

The supersymmetric sigma model on $CP^1=S^2$ is a massive $N$=2 theory.  In
principle we can consider an arbitrary kahler metric on $CP^1$. Varying the
metric without changing its kahler class (i.e., preserving the area) is a
$D$-term perturbation of the theory and therefore the index only depends on
the area.  Letting $X$ denote the Kahler form of $CP^1$, the
chiral ring, which is to be identified as
an instanton-modified cohomology ring, is $X^2=e^{-A}$ where $A$ is the area of
$CP^1$ (the two-sphere) \rNar .  The $tt^*$ considerations have been applied
to this and $CP^{n-1}$ examples in \rCVii : using the above ring the index is
again given by \piii\ where $z=8\beta e^{-A/2}$
and, by regularity, the boundary condition
\piiibc\ is determined to be $\Theta =0$, i.e. $r$=2.

For a generic metric on $CP^1$, the supersymmetric sigma model is not an
integrable theory. However, since the index only depends on the area, for
computation of the index we may as well use a convenient choice of the metric.
If we choose the constant-curvature metric on $CP^1$ then this theory is
integrable \intcpnrefs\ and so we can compare the $tt^*$ analysis with the TBA
analysis.  As discussed in \pkii , the TBA system for this theory is obtained
by taking the $k\rightarrow \infty$ limit of the Chebyshev TBA system.  Our
index for the supersymmetric $CP^1$ sigma model is thus given by \pintt\ and
\piiiFt\ with $\Theta =0$, in agreement with the $tt^*$ result, given our
equality between the Painleve III differential equations and these integral
equations.

\subsec{The Basic ${\bf Z}_n$-type $N$=2 Integrable Theories}

All of the previous examples displayed a spontaneously broken ${\bf Z}_2$
symmetry.  The $tt^*$ equations for the index in all these examples were the
same, the only difference being in the boundary conditions.  Likewise, the TBA
integral equations for the index in all of these examples were the same, the
only difference being in the value of $\Theta$ in \pintt .  We will now
consider $N$=2 integrable theories with a spontaneously broken ${\bf Z}_n$
symmetry.  The basic such theory is the $A_n$ $N$=2 minimal conformal field
theory perturbed by the most-relevant supersymmetry preserving operator.  It
can be described by the Landau-Ginzburg superpotential \rVW
\eqn\mrp{W=\lambda ({X^{n+1}\over n+1}-X ).}
This theory is integrable \LGSol; in fact, it can be described by an affine
Toda theory with an imaginary coupling and a background charge \toda.

The $tt^*$ equations for this example were discussed in \cv .  Because theory
\mrp\ has the ${\bf Z}_n$ symmetry $X\rightarrow e^{2\pi i/n}X$, the Ramond
ground state metric $g$ is diagonal in the chiral ring basis spanned by $1, X,
\dots , X^{n-1}$.  Denoting these diagonal elements by $e^{q_p}$, equation
\curv\ with the chiral ring relation $X^n$=1 yields the following relations
for the eigenvalues $Q(z;p)$ of our matrix index:
\eqn\toda{Q(z;p)=z{d\over dz}q_p(z)\quad \hbox{where}\
{d^2q_p\over dz^2}+{1\over z}{dq_p\over dz}+e^{q_{p+1}-q_p}-e^{q_p-q_{p-1}}=0,}
for $p=0, \dots , n-1$, with $q_{p+n}\equiv q_p$ and with the constraint
$q_{n-p}=-q_{p}$.  For $n$=2, \toda\ reduces to the sinh-Gordon \piii\ and
\piiiQ\ of our previous examples.  The ${\hat A}_{n-1}$ Toda
equations with the constraint $q_{n-p}=-q_{p}$ can be put in the form of
${\hat C}_m$ Toda theory for $n=2m$ or ${\hat {BC}}_m$ Toda theory for
$n=2m+1$ \cv .

Exact integral equations for the above eigenvalues of our matrix index follow
from the exact $S$-matrix and TBA analysis discussed in \pkii .  The vacua of
theory \mrp\ are at $X^{(j)}=e^{2\pi ij/n}$. The soliton content consists of
$2(n-1)$ solitons forming doublets $(u_r, d_r)$ under supersymmetry; the
soliton species label $r$ runs from $1, \dots ,n-1$ corresponding to solitons
connecting initial and final vacua with $X_f=e^{2\pi ir/n}X_i$.  The fermion
numbers of $(u_r, d_r)$ are given by $(r/n,r/n-1)$, and their mass is given by
the Bogomolnyi bound $m_r=M\sin r\pi /n$ where $M=4n\lambda /(n+1)$.

The conserved currents require that when a soliton of type $r$, i.e. $u_r$ or
$d_r$, scatters with a soliton of type $s$, the labels $r$ and $s$ scatter
diagonally --- along with the rapidities. The number $N_r$ of solitons of type
$r$, i.e. the number of $u_r$ solitons plus the number of $d_r$ solitons, is
thus conserved for $r=1, \dots ,n-1$, as is the total number of $u$ solitons,
and the total number of $d$ solitons.  A multi- soliton state can thus be
characterized by distributions $\rho _r (\t )$ of rapidities occupied by
solitons of type $r$, i.e. $u_r$ or $d_r$, along with, as in the previous
examples, two additional distributions $\rho _l(\t )$ and $l=0, \bar 0$ \pkii.
The fermion number of the multi-soliton state is again found to be given in
terms of the various distributions by \fis .

Using the exact $S$-matrix it was found in
\pkii\ that the distributions $P_a(\t)$ of
available levels for the above species $a$
are given in terms of the above occupied
distributions $\rho _a(\t )$ by relations of the usual form
\Pdist\ where $m_r=M\sin (r\pi/n)$ is the mass of species $r=1, \dots ,n-1$,
$m_l=0$ for $l=0, \bar 0$, and the $\phi _{ab}$ are given by
\eqn\phiab{\eqalign{\phi _{l,l'}&=0, \qquad
\phi _{r,l}(\t )={\sin (r\mu)\over
\cosh (\t )- a_l\cos (r\mu )},\cr
\phi_{r,s}(\t)&=\int dt e^{it\t}
\left(\delta_{rs}-2{\cosh \mu t \sinh(\pi-r\mu)t \sinh s\mu t \over \sinh \pi
t\sinh \mu t}\right),\cr}}
for $r\geq s=1, \dots n-1$, with $\phi _{ab}=\phi _{ba}$, where $a_0\equiv -
a_{\bar 0}\equiv 1$ and $\mu \equiv \pi/n$.

We now calculate the eigenvalues of $\Tr _{ab}(e^{i\alpha F}e^{-\beta H})$.
The $p$-th eigenvalue of this matrix can be obtained by weighting solitons of
type $r$ by $e^{2\pi irp/n}$, for $p=0, \dots , n-1$, and summing over all
vacua $(ab)$ at spatial infinity.  We will thus calculate $\log \Tr e^{i\alpha
F}e^{2\pi i prN_r/n}e^{-\beta H}$ by using \Emu\ with the chemical potentials
$\beta \mu _r=-2\pi irp/n$ and, from \fis , $\beta \mu _{\bar 0}=-\beta \mu
_0=i\alpha$.  Plugging these chemical potentials and the \phiab\ into \TBA\
and \fe , we obtain integral equations which determine exactly $\Tr(e^{i\alpha
F}e^{-\beta H})$.

At $\alpha=\pi$ it is seen by inspection that the solution of the coupled
integral equations \TBA\ is given by $e^{-\e_r(\tht)}=\e _l(\tht)=0$, and thus
$L^{-1}\log \Tr (-1)^Fe^{-\beta H}=0$ in the large $L$ limit.
We now consider $\alpha=\pi +h$ with $h$ small.  The solution
of \TBA\ is of the form $e^{-\e _r}=he^{-2\pi i rp/n} e^{-A_r}$ and $\e _l=h
B_l$, where the $A_r$ and $B_l$ satisfy the coupled integral equations
\eqn\AB{\eqalign{A_r(\t )&=m_r\beta \cosh \t +{2 irp\mu}
-\sum _{l=0, \bar 0}\int {d\t'\over 2\pi}{\sin (r\mu )
\ln (ia_l+B_l (\t '))\over \cosh (\t -\t')-a_l\cos (r\mu )}\cr
B_l(\t ')&=-\sum _{r=1}^{n-1}\int {d\t ''\over 2\pi}{\sin (r\mu)e^{-A_r(\t
'')}\over \cosh (\t ' -\t '')-a_l\cos (r\mu )},\cr}}
where, again, $a_0=-a_{\bar 0}=1$,  $m_r=M\sin (r\pi /n)$, and $\mu=\pi /n$.
Using \indd\ and
\fe\ we have the index eigenvalues $Q(z=M\beta ;p)$, for $p=0, \dots ,n-1$,
given by
\eqn\theind{Q(z;p)=\sum _{r=1}^{n-1} m_r\beta \int {d\t \over
2\pi}\cosh \t e^{-A_r(\t ;z;p)}.}
It follows from \AB\ that $A_{r}(\t ;z;p)^*=A_{n-r}(\t ;z;p)$ and $B_0(\t
;z;p)^*=B_{\bar 0}(\t ;z;p)$ and, thus, the above eigenvalues $Q(z;p)$ are all
real, as they should be.  Furthermore, $e^{-A_r(\t ;z;p)}=-e^{-A_{n-r}(\t
;z;n-1-p)}$ is a consistent solution of \AB\ for all $r=1, \dots ,n-1$ and
$p=0, \dots, n-1$.  It follows that $Q(z;n-1-p)=-Q(z;p)$; the non-zero
eigenvalues come in opposite pairs as they should.

We can compare these exact results \theind\ and \AB\ with the infra-red
expansion of sect.\ 4.  Setting $B_l(\t ')=0$ in \AB , we obtain the first
approximation $e^{-A_r}\approx ie^{-i\pi r(2p+1)/n}e^{-m_r\beta \cosh \t}$.
Plugging this into
\theind\ we obtain the one-soliton sector contribution to the index, in
agreement with \ipart.
Plugging this first approximation back into \AB\ we obtain the next
approximation
$$\eqalign{e^{-A_r}\approx ie^{-i\pi r(2p+1)/n}&e^{-m_r\cosh \t }\cr
&(1-\sum _se^{-i\pi s(2p+1)/n}\int
{d\t '\over 2\pi}\phi_{r+s,0}(\t -\t' )e^{-m_s\beta \cosh \t '}).\cr}$$
Plugging this into \theind\ gives the contribution to the index coming from
the two-soliton sector, in agreement with \Qiisimp\ and \aatilsat .

We have again found two different representations of our index:  One in
terms of solutions to affine Toda equations \toda\ and the other in
terms of solutions to coupled integral equations \AB\ and \theind .
It would be interesting to check this equivalence numerically and
verify it analytically.

\subsec{Other ${\bf Z}_n$ Integrable Theories}

There are a variety of other integrable $N$=2 theories with
spontaneously-broken ${\bf Z}_n$ symmetry whose index can be obtained from the
equations of the previous subsection.  Examples are the affine Toda
generalizations of $N$=2 sine-Gordon, integrable perturbations of $N$=2
Kazama-Suzuki theories described by the $SU(n)_k$ generalized Chebyshev
polynomial superpotentials in $n-1$ variables \refs{\rGep} and the $CP^{n-1}$
sigma models \pkii .  We first consider the $SU(n)$ affine Toda theories
described by the action
\eqn\ata{\int d^2zd^4\t \sum _{j=1}^{n-1}X_j\bar X_j+{M\over
4n}(\int d^2zd^2\t
\sum _{j=1}^{n}e^{ig(X_j-X_{j-1})}+h.c.),}
where $X_0\equiv X_n\equiv 0$.  The vacua of this theory form the $n-1$
dimensional weight lattice of $SU(n)$.  This theory has $n-1$ topological
charges $T_r$ and, as in the sine-Gordon case, the eigenvalues of the $Q$
matrix index can be written as
\eqn\QTgen{Q(z;\Theta _1, \dots ,\Theta _{n-1})=i\beta L^{-1}\Tr F(-1)^F
e^{i\sum_{r=1}^{n-1}\Theta _rT_r}e^{-\beta H},}
where the trace runs over all boundary conditions.

The $tt^*$ analysis has been applied to this example in \cv\ where
the solutions are expressed in terms of solutions to the affine
toda equations \toda\ but now with different boundary conditions,
which should now depend on $\Theta_i$.

The $S$-matrix for the theory \ata\ is (conjectured
to be) the tensor product of
the $N$=2 theory discussed in the previous subsection with an additional $N$=0
structure with vacua corresponding to the weight lattice of $SU(n)$: vacua
labelled by $SU(n)$ weights $\mu$ and $\nu$ are connected by a soliton doublet
of the $r$-th type, with fermion numbers $(r/n,r/n-1)$ and mass $m_r=M\sin
(r\pi /n)$, provided the representations satisfy $\mu\otimes \Lambda _r =\nu
\oplus \dots $ where $\Lambda _r$ is the
$r$-th fundamental representation of $SU(n)$
($r=1, \dots ,n-1$).  As far as $N$=2 supersymmetry is concerned, every
$r$-type soliton doublet is identical to the $r$-type doublet in the basic
\Zn\ theory discussed in the previous section.  We will not explicitly carry
out the TBA analysis for this theory.  Rather, we will use the fact that, as
seen in the previous examples, the only effect of the additional $N$=0
structure on our index is to modify the $A_r(\t )$ equations in \AB\ with some
$\t$-independent constants $C_r(\Theta _j)$ (generalizing
the term $2\cos \Theta$ in \pintt\ for the affine toda case):
\eqn\Qgen{Q(z;\Theta _1, \dots , \Theta _{n-1})=\sum
_{r=1}^{n-1} m_r\beta \int {d\t \over
2\pi}\cosh \t e^{-A_r(\t )},}
where the $A_r(\t )$ are solutions to the coupled integral equations
\eqn\ABgen{\eqalign{A_r(\t )&=m_r\beta \cosh \t -\ln C_r(\Theta _j)
-\sum _{l=0, \bar 0}\int  {d\t'\over 2\pi}{\sin (r\mu )
\ln (ia_l+B_l (\t '))\over \cosh (\t -\t')-a_l\cos (r\mu )}\cr
B_l(\t ')&=-\sum _{r=1}^{n-1}\int {d\t ''\over 2\pi}{\sin (r\mu)  e^{-A_r(\t
'')}\over \cosh (\t ' -\t '')-a_l\cos (r\mu )}.\cr}}
These equations lead to an expansion of the form
\eqn\solexp{Q(z; \Theta _j)=\sum _{N_1 =1}^\infty \cdots
\sum _{N_{n-1}=1}^\infty
C_1^{N_1}\cdots C_{n-1}^{N_{n-1}}A_{N_1, \dots ,N_{n-1}}(z),}
where $A_{N_1, \dots N_{n-1}}(z)$ is the contribution from the sector with
$N_r$ solitons of type $r$ for $r=1, \dots , n-1$.  Comparing \solexp\ with
\QTgen\ it is seen that the $C_r(\Theta _j)$ are given by the character
functions
$$C_r(\Theta _j)=\sum _{\lambda \in L(\Lambda _r)}e^{i\lambda \cdot T},$$
where $L(\Lambda _r)$ are the weights in the $r$-th
fundamental representation $\Lambda_r$ of $SU(n)$ (the $SU(n)$ representation
with $r$ vertical boxes for its Young tableau), for $r=1, \dots ,n-1$, and
$T=\sum _n\Theta _n\alpha _n$ where $\alpha _n$ are the simple roots of
$SU(n)$.  The above characters can be written (as in \rGep ) as
\eqn\zT{\sum _{l=0}^nC_l(\Theta _1, \dots , \Theta _{n-1})t^l=\prod
_{j=1}^n (1+te^{i(\Theta _j-\Theta _{j-1})}),}
where $\Theta _0\equiv \Theta _n\equiv 0$; expanding the product in $t$ and
equating coefficients of $t^l$ on both sides yields the above sums over the
fully antisymmetric representations.  Plugging these $C_r(\Theta _i)$ into
\ABgen\ and \Qgen\ yields the index $Q(z; \Theta _j)$.  It follows from \zT\
that $C_r(\Theta _i)^*=C_{n-r}(\Theta _i)$ from which it follows from \ABgen\
that $A_r(\t ;z;\Theta _i)^*=A_{n-r}(\t ;z;\Theta _i)$ and $B_0(\t ;z;\Theta
_i)^*=B_{\bar 0}(\t;z;\Theta _i)$.  It then follows that the index eigenvalues
$Q(z;\Theta _i)$ are real.

The above integral equations must provide the regular solutions of the Toda
differential equations \toda .  As in the PIII case, physics has proven a
statement for which there is, as of yet, no purely mathematical proof.
In particular this gives an $n-1$ parameter family of regular solutions
to radial affine Toda equations \toda .
The dependence of the solutions on these
parameters ($\Theta_i$) is again in line with the intuition
discussed in the case of PIII (see footnote 12).

We now consider the theory with superpotential given by the $SU(n)_k$
Chebyshev polynomial $W_{n+k}(X_1, \dots ,X_{n-1})$ in $n-1$ variables
discussed in \rGep ; the generating function for these potentials is
\eqn\Wink{\sum _pW_p(X_1, \dots ,X_{n-1})t^p=-\log
(1+\sum _{r=1}^{n-1}X_r(-t)^r+(-t)^n).}
These theories have been discussed in \refs{\rGep , \cv , \rken , \nw ,\pkii
,\rGepii} ($Sp(N)_k$ theories, which might also be integrable, have been found
in \refs{\Bord ,\rGepii}).  {}From the $tt^*$ analysis it is possible to see
that the index is again related to the affine toda equation \cv .
{}From the TBA integral equations the
$(n+k-1)!/(n-1)!k!$ eigenvalues of the matrix
index for the $SU(n)_k$ theory are expected to be
obtained from the equations \Qgen\ and
\ABgen\ by setting $C_r (\Theta _i)=X_r(\mu )$ where $X_r(\mu )$ are the
solutions of $dW_{n+k}(X_r)=0$.  This is equivalent \rGep\ to setting
$C_r(\Theta _i)=S_{\Lambda _r \mu}/S_{0\mu}$, where $S_{\mu \nu}$ is the
$SU(n)_k$ modular transformation matrix, $\Lambda _r$ is the $r$-th
fundamental representation $SU(n)$, $\mu$ is one of the $(n+k-1)!/(n-1)!k!$
highest weight representations of $SU(n)_k$, and $0$ is the
identity.  These integral equations for the index eigenvalues of the
$SU(n)_k$ Chebyshev theory generalize the $SU(2)_k$ results in subsect. 6.3
and the $SU(n)_1$ results in
sect. 6.5.
Finally, we consider the $CP^{n-1}$ sigma model.  Here again, the
$tt^*$ equations give the affine
toda equations \rCVii\ (with logarithmic boundary conditions).
The $S$-matrix
for $CP^{n-1}$ sigma model is obtained from the $k\rightarrow
\infty$ limit of this $SU(n)_k$ Chebyshev theory \pkii .  Thus, the index for
this sigma model is also given by the above integral
equations by setting the $\Theta _r=0$, i.e. $C_r=n!/r!(n-r)!$, in \ABgen .

\newsec{Conclusions}
We have seen that for two-dimensional $N$=2 supersymmetric theories ${\rm
Tr}(-1)^F F e^{-\beta H}$ is a (matrix) index in a generalized sense, i.e., it
is independent of $D-$term perturbations.  Though the index is not
topological, it is determined exactly via non-linear differential equations
which are obtained using only topological data, namely the chiral ring.  These
non-linear differential equations encode the geometry of the vacua of the
theory.  This allows us to read off, by an IR expansion, the spectrum of
Bogomolnyi saturated states of the theory as well as some aspects of their
interaction.  In case the theory is integrable and the exact $S$-matrix is
known, the index can be computed using TBA methods in terms of solutions to
coupled integral equations.  It is a very non-trivial statement that in these
cases the integral equations thus obtained are equivalent to the differential
equations characterizing the geometry of vacua.

It is amusing that one can apply $N$=2 formalism to study polymer physics \fs.
The index in this context is the partition function of a single polymer
wrapped around a cylinder of perimeter $\beta$.

Given the fact that the integral equations which arose for us in the
context of the TBA are equivalent to differential equations, it is very
natural to ask if this can be done more generally.  In other words,
is it {\it always} possible to relate integral equations arising
for integrable theories through the TBA to ordinary differential equations?
A first step in this direction may be to try to prove mathematically
why in our case the integral equations of the TBA were equivalent to
differential equations.

For non-integrable theories the $tt^*$ equations can still be used to compute
the index.  In the infra-red the leading contribution to the index is
universal \ipart .  However, even though the normalization of this term can be
easily deduced from a Hilbert space interpretation, from the viewpoint of
solving the $tt^*$ equations this
is only fixed by requiring the regularity of the
solution (even in the UV regime).  Therefore, it is a very non-trivial test of
these ideas that the normalization coming from solving the $tt^*$ equations in
the IR agrees with the Hilbert space interpretation. This has recently been
confirmed even in a non-integrable case by solving $tt^*$ equations
numerically \bill .

We have seen that the index basically captures the geometry and interaction of
kinks interpolating between supersymmetric vacua.  It would be interesting to
write a general solution (say for Landau-Ginzburg theories) of the $tt^*$
equations in terms of these kinks.  Such a thing is not unexpected, given the
fact that $tt^*$ equations depend only on the superpotential (which is
equivalent to knowing the kink spectrum and their geometry) and that the
equations are integrable as they can be rephrased as flatness conditions even
if the underlying quantum field theory is not integrable, a fact which has
been recently elaborated upon in \dub .

 Given the power of the new supersymmetric index in encoding exact results, it
would be tempting to look for similar objects in other supersymmetric
theories.  In particular a very similar setup to what we have discussed in
this paper appears naturally in the context of four-dimensional $N$=2
Yang-Mills theory.  Again this theory is related to a topological theory
\Witd\ and the analog of the chiral fields are the two-cycle observables.  In
particular for the $SU(2)$ gauge theory in the Higgs phase where $SU(2)$ is
broken to $U(1)$, all the known particles of the theory such as the massive
gauge particles and the monopoles are known to saturate the Bogomolnyi bound
\rWO\ very much as our kinks in the two-dimensional theory saturate the
Bogomolnyi bound.  In this case the natural generalization of our index seems
to be ${\rm Tr}(-1)^F J^2 e^{-\beta H}$ where $J$ is the generator of $U(2)$
symmetry of $N$=2 theories\ref\cvp{S.  Cecotti and C. Vafa, work in
progress.}. It would be exciting to see what exact information about the
$S$-matrix of these four-dimensional theories are encoded in such an index.

\nref\dix{L.J. Dixon, V.S. Kaplunovsky and J. Louis,
Nucl. Phys. B 355 (1991) 649.}
\nref\narra{I.Antoniadis, E. Gava and K.S. Narain,
preprints IC/92/50; IC/92/51.}
To formulate our index we needed to put the
Hilbert space on infinite line to allow for kinks.  If we put the
Hilbert space on a periodic circle and thus
compute the index on the torus we get
zero.  This can also be seen by CPT invariance.  However if we replace $F$ by
$F_L$, the left-moving fermion number, in the definition of the index, CPT no
longer requires it to vanish (as even on the torus $(-1)^{F_L}$ is in general
not $\pm 1$).  This quantity has already appeared in the context of conformal
theories where it is related to the moduli dependence of the gauge and
gravitational coupling constants \refs{\dix,\narra}.  This modified index
resembles the generalization of Ray-Singer torsion
\ref\sing{D.B. Ray and I.M. Singer, Adv. Math. 7 (1971) 145;
Ann. Math. 98 (1973) 154}\ to conformal theories.

We have seen that $N$-kink configurations each contribute to our index through
an `anomaly' resulting from an inequality in the density of states for a
supersymmetric multiplet \ivdiml.  Each of these contributions reminds one of
(though it is not the same as) the Callias-Bott-Seeley index \cab. It
would be very exciting to uncover the meaning of such a `topological
invariant' for each $N$-kink contribution.  Our new index, which sums up the
contribution of all $N$-kink configurations, would then encode infinitely many
topological invariants into a single function!

\centerline{\bf Acknowledgements}

We would like to thank C.Imbimbo, W. Leaf-Herrmann, A. Lesniewski and E.
Witten for helpful discussions.  C.V. would also like to thank the ICTP, where
part of this work was done, for hospitality.  P.F.\ was supported by DOE grant
DEAC02-89ER-40509, K.I.\ was supported by NSF grant PHY-87-14654 and an NSF
graduate fellowship and C.V.\ was supported by Packard fellowship and NSF
grants PHY-87-14654 and PHY-89-57162.

\appendix{A}{An Operatorial Proof of the tt* Equations}

In this appendix we show how the ideas of the present paper can be used to
give a quick (although less rigorous) proof of the $tt^*$ equations of \cv .

We begin by rewriting the $tt^*$ equations in terms of the $Q$ matrix only. We
have seen in the main body of the paper that
\eqn\known{Q_{a b}={i\beta\over L} {\rm Tr}_{a b} (-1)^F F
e^{-\beta H},}
where $a$, $b$ label  boundary conditions at spatial
infinity associated to some basis $|a\rangle$ of vacua.

In terms of $Q$ the basic $tt^*$ equations read
\eqn\oldstuff{\eqalign{&\bar D_{\bar i} Q=
2\beta^2 [C_\tau, \bar C_{\bar i}]\cr
& D_{i} Q
=2 \beta^2[C_i, \bar C_{\bar\tau}].\cr} }
\eqn\dimensions{\eqalign{&\half [Q,C_j]=-C_j+D_j C_\tau\cr
&\half [Q,\bar C_{\bar j}]=\bar C_{\bar j}-\bar D_{\bar j}\bar C_{\bar
\tau}.\cr} }
where $D_i$ ($\bar D_{\bar i}$) denotes the metric covariant derivative and
$C_\tau$ is the matrix representing multiplication by $W$ (in the case of a
Landau-Ginzburg theory, which we assume in this section for simplicity) on the
vacua $|a\rangle$.  One can think of $Q$ in the special parametrization of the
couplings of the theory motivated from the
renormalization group as $Q=2(A_\tau
-A_{\tau^*})$ with $\beta = \exp(\tau /2+ \tau^* /2 )$, from which one can
deduce the above equations from the usual $tt^*$ form (note that a chiral
operator has an explicit $\tau$ dependence given by $e^\tau \phi_i$).  The
above equations are written so as to make sense in an arbitrary basis (see
also \cec ).  Eq.\dimensions\ has the following interpretation.  For a
quasihomogeneous family of superpotentials $C_\tau=0$ and
\dimensions\ just states that the marginal chiral (resp. anti-chiral)
deformations have $U(1)$ charge $+1$ (resp. $-1$). The extra term in the
 RHS  measures the `deviation from marginality'.

On the other hand, \dimensions\ allows us to write the full metric connection
$A_i$ in terms of $Q$ and the ring coefficients. Then we can compute its
curvature in terms of $Q$.  Because of this, \oldstuff\ and \dimensions\
together with
\eqn\othereqs{\eqalign{&D_i\bar C_{\bar j}=\bar D_{\bar j}C_i=0\cr
& D_iC_j=D_jC_i,\cr} }
reproduce all the tt* equations. (In fact, the second line is a consequence
of \dimensions\ together with known properties of the chiral ring).
In particular, we get
$$\eqalign{&[D_i,\bar D_{\bar j}]Q=-\beta^2[[C_i,\bar C_{\bar j}], Q]\cr
&[D_i,\bar D_{\bar j}]C_{\tau}=-\beta^2[[C_i,\bar C_{\bar j}], C_\tau]\cr}$$
from which we read the curvature of the metric connection\foot{At least for a
generic superpotential, these equations fix the curvature
un\-am\-big\-uously.}.

One has also the identity ($\Delta$ is defined in \sual )
\eqn\central{[C_\tau, \bar C_{\bar j}]_{a b}=\half {\rm
Tr}_{a b}(-1)^F \Delta \bar\phi_j e^{-\beta H}.}
This equation deserves a comment. The simplest way of proving it is to
choose the vacuum basis $|a\rangle$ to correspond to the canonical
basis, i.e. the holomorphic point basis (normalized so that $\eta=1$).
Then the central charge has a definite value $\Delta_{ab}=2[W(a)-W(b)]$, and
\central\ follows from the definition of $C_\tau$ and the obvious identity
\eqn\cibar{(\bar C_{\bar i})_{a b}=
 {\rm Tr}_{a b} (-1)^F \bar\phi_{\bar i} e^{-\beta H}. }
However, the canonical basis is not the natural one from a `thermodynamical'
viewpoint. In this framework one decomposes the Hilbert space $\cal H$ into
sectors for which $\Delta$ {\it and} $\bar \Delta$ have a definite value. Such
sectors should exist on general grounds. Now, whereas it is manifest that the
canonical boundary conditions give a definite value for $\Delta$, it seems
unlikely that they also have a definite $\bar\Delta$.  Roughly speaking, the
natural boundary conditions should correspond to the `real' point basis for
vacua, defined by prescribing the asymptotical value of the scalar fields to
to be a classical vacuum\foot{ This can be made more precise by defining the
`real' point vacua by starting from a large circle to quantize the theory
where the point basis is unambiguous and adiabatically change the radius of
the circle.}.  We have two comments: First the identity above, being covariant
under changes of bases, should be valid even in such a `real' basis. Second, in
the simplest situations we can explicitly construct the `real' point basis and
check the consistency of our formal manipulations.  At any rate it would be
worthwhile understanding the real point basis more clearly.

The new proof of tt* consists in showing that Eq.\oldstuff\ and \dimensions\
follows from the representation \known\ of $Q$ and the `AB argument'.  At the
formal level we have (using the `AB argument')
$$\eqalign{\delta_i & {\rm Tr}_\ast (-1)^F F e^{-\beta H}=\cr
&=i\beta L {\rm Tr}_\ast (-1)^F F\{Q^-,[\bar Q^-,\phi_i]\}e^{-\beta H}\cr
&=-i\beta L {\rm Tr}_\ast (-1)^F \{Q^-,\bar Q^-\}\phi_i e^{-\beta H}\cr
&=-i\beta L {\rm Tr}_\ast (-1)^F \bar \Delta \phi_i e^{-\beta H},\cr}$$
where $\ast$ means some sector $(a,b)$ of the Hilbert space. Clearly, in view
of \known\ and \central , this is the same as \oldstuff\ provided we
interpret $\delta_i$ as $D_i$, i.e. as the metric covariant derivative. This
is the correct interpretation.  In general we get some contribution to the
derivative of ${\rm Tr}_\ast (-1)^F F \exp[-\beta H]$ from the variation of
the boundary condition $\ast$. Such terms have a structure which allows to
absorb them in the definition of the connection in $D_i$. This is natural,
because in a sense the path integral variation should `dress' the vacua at
infinity as well to make them be the new vacua.  In this interpretation, for
example, what the `AB argument' for invariance of the Witten's index discussed
in section 2 really shows is that ground state metric $g$ is covariantly
constant.  Similarly here this suggests that we have {\it some} covariant
derivative such that
$$D_i {\rm Tr}_\ast (-1)^F F e^{-\beta H}=-i\beta L {\rm Tr}_\ast
(-1)^F\bar\Delta\phi_i e^{-\beta H}.$$
(And analogously for $\bar D_{\bar i}$). The connection cannot be trivial.
Indeed, as shown above, we can use the resulting equation to compute its
curvature which turns out to be non--vanishing.  It remains to show that the
connection predicted by this argument is the metric one. Indeed the AB
argument predicts
$$D_i {\rm Tr}_\ast (-1)^F e^{-\beta H}=\bar D_{\bar i}{\rm Tr}_\ast (-1)^F
e^{-\beta H}=0,$$
for the  connection induced by the variation of the boundary condition $\ast$.

 The same reasoning applied to \cibar\ gives
$$D_i (\bar C_{\bar j})_\ast= D_i {\rm Tr}_\ast (-1)^F \bar\phi_{\bar
j}e^{-\beta H}= -i\beta L {\rm Tr}_\ast (-1)^F\bar\phi_{\bar j}
 \{Q^-,[\bar Q^-,\phi_i]\}e^{-\beta H}=0$$
$$D_i(C_j)_\ast=D_i{\rm Tr}_\ast (-1)^F \phi_j e^{-\beta H}=-i{\rm Tr}_\ast
(-1)^F \phi_j \{Q^-,[\bar Q^-,\phi_i]\}e^{-\beta H}= D_j(C_i)_\ast,$$
showing \othereqs .

Finally we show \dimensions . By definition, one has (as $L\rightarrow
+\infty$)
$$\eqalign{[Q,C_j]_\ast &= {i\beta\over L}{\rm Tr}_\ast (-1)^F F
e^{-\beta H} \Big[\phi_j(x=+L/2)-\phi_j(x=-L/2)\Big] \cr
&= {i\beta\over L}{\rm Tr}_\ast (-1)^F F  e^{-\beta H}
\int_{-L/2}^{+L/2} d_x\phi_j.\cr}$$
For $L$ large, we can replace
$$\int_{-L/2}^{+L/2}d_x\phi_j\rightarrow {iL\over 2}\{Q^+,[Q^-,\phi_j]\}-
{iL\over 2}\{\bar Q^+,[\bar Q^-,\phi_j]\}.$$
In this way we get an expression to which we can apply the `AB argument'.
One gets
$$\eqalign{[Q, C_j]_\ast &={i\beta\over 2}{\rm Tr}_\ast (-1)^F
 e^{-\beta H}\big[\{Q^+,Q^-\}+
\{\bar Q^+,\bar Q^-\}\big] \phi_j \cr
&=- \beta  {\rm Tr}_\ast (-1)^F   e^{-\beta H}H \phi_j
= \beta D_\beta\left[{\rm
Tr}_\ast (-1)^F \phi_j e^{-\beta H}\right].\cr}$$
In view of \othereqs\ and dimensional analysis (i.e. dependence of the fields
on the scale) this equation is equivalent to \dimensions . Indeed, this is the
covariant version of the statement (true only in special `gauges') that $Q$ is
the connection in the $\tau$ direction. To compare with computations done in
such special gauges recall that $\beta =e^{\tau /2+\tau^* /2}$ and so $\beta
g\partial_\beta g^{-1}= 2g\partial_\tau g^{-1}$.

\listrefs

\end